\documentclass{aastex631}
\usepackage{mathrsfs}
\usepackage{color}
\usepackage{comment}
\usepackage{amsmath}
\usepackage{amssymb}
\def\be{\begin{equation}}
\def\ee{\end{equation}}
\def\ba{\begin{eqnarray}}
\def\ea{\end{eqnarray}}
\def\msun{M_\odot}

\def\Omvec{{\mbox{\boldmath $\Omega$}}}

\def\muhat{{\mbox{\boldmath ${\hat\mu}$}}}
\def\muvec{{\mbox{\boldmath $\mu$}}}

\def\Omhat{{\mbox{\boldmath ${\hat\Omega}$}}}
\def\Ivec{{\mbox{\boldmath $I$}}}
\def\Lvec{{\mbox{\boldmath $L$}}}
\def\Nvec{{\mbox{\boldmath $N$}}}

\def\evec{{\mbox{\boldmath${\hat e}$}}}

\def\Bvec{{\mbox{\boldmath$B$}}}

\def\rvec{{\mbox{\boldmath$r$}}}
\def\xvec{{\mbox{\boldmath$x$}}}

\def\dotprod{{\mbox{\boldmath$\cdot$}}}
\def\crossprod{{\mbox{\boldmath$\times$}}}
\def\rhat{{\mbox{\boldmath${\hat r}$}}}
\def\phihat{{\mbox{\boldmath${\hat\varphi}$}}}

\def\ehat{\evec}
\def\nhat{{\mbox{\boldmath${\hat n}$}}}
\def\bhat{{\mbox{\boldmath${\hat b}$}}}

\def\kvec{{\mbox{\boldmath${k}$}}}

\def\Lvec{{\mbox{\boldmath$L$}}}

\def\lvh{{\mbox{\boldmath${\hat\ell}$}}}

\def\MeV{{\rm MeV}}
\def\keV{{\rm keV}}
\def\Gauss{{\rm Gauss}}
\def\Oscr{{\mathscr{O}}}

\def\uhat{{\mbox{\boldmath${\hat u}$}}}

\def\grad{{\mbox{\boldmath$\nabla$}}}
\def\curl{\grad\crossprod}
\def\costheta{\cos\theta}

\def\gram{{\rm g}}
\def\erg{{\rm erg}}
\def\cm{{\rm cm}}
\def\nnuc{n_{\rm nuc}}
\def\fermi{{\rm fm}}

\def\degK{\,{\rm K}}

\def\hctwo{H_{{\rm c}2}}

\def\epsrot{\epsilon_{\rm rot}}

\def\Emag{E_{\rm mag}}
\def\epsmag{\epsilon_{\rm mag}}
\def\epsmagT{\epsilon_{{\rm mag},T}}
\def\epsmagD{\epsilon_{{\rm mag},D}}

\def\eps{\epsilon}
\def\deps{\Delta\eps}

\def\Iunit{I_{0,45}}

\def\sec{{\rm s}}

\def\dstar{d^\star}

\def\epsmag{\eps_{\rm mag}}

\def\tsd{t_{\rm sd}}

\def\zhat{\ehat_z}

\def\costheta{\cos\theta}

\def\cn{\,{\rm cn}}
\def\sn{\,{\rm sn}}
\def\dn{\,{\rm dn}}
\def\cosi{\cos i}
\def\sini{\sin i}

\def\pfp{p_{\rm F,p}}
\def\Psec{P({\rm s})}
\def\Phipcyc{{\Phi_{p,{\rm cycle}}}}
\def\Phicyc{\Phipcyc}
\def\phicyc{\phi_{p,{\rm cyc}}}

\def\Ppdays{P_p({\rm d})}

\def\snsqav{\langle\sn^2\Phi\rangle}
\def\cnsqav{\langle\cn^2\Phi\rangle}
\def\dnsqav{\langle\dn^2\Phi\rangle}
\def\Dt{\Delta t}

\def\epst{\epsilon t}
\def\Phitil{{\tilde\Phi}}

\def\DIvec{\Delta\Ivec}
\def\DI{\Delta I}

\def\fvec{\mbox{\boldmath{$f$}}}
\def\zhat{\mbox{\boldmath{${\hat z}$}}}
\def\km{{\rm km}}
\def\that{\mbox{\boldmath{${\hat t}$}}}
\def\vhat{\mbox{\boldmath{${\hat v}$}}}

\def\asin{{\sin^{-1}}}

\def\dPsi{{\delta\Psi}}

\def\btms{\langle B_T^2\rangle}
\def\bturb{B_{\rm turb}}

\def\nh{{\hat n}}
\def\Dhat{{\hat D}}
\def\nustar{\nu_\star}

\def\epsd{{\epsilon_{\rm sd}}}

\def\tnh{\theta_{n}}
\def\phinh{\varphi_{n}}
\def\ntil{{\tilde n}}
\begin{document}

\title{Non-Axisymmetric Precession of Magnetars and Fast Radio Bursts}
\author{I. Wasserman}
\affiliation{Cornell Center for Astrophysics and Planetary Science, Cornell University}
\affiliation{Laboratory for Elementary Particle Physics, Cornell University}
\correspondingauthor{I. Wasserman}
\email{ira@astro.cornell.edu}
\author[0000-0002-4049-1882]{J. M. Cordes}
\author[0000-0002-2878-1502]{S. Chatterjee}
\author{G. Batra}
\affiliation{Cornell Center for Astrophysics and Planetary Science, Cornell University}


\keywords{stars: neutron --- stars: magnetars --- Fast Radio Bursts: FRB 121102  --- Fast Radio Bursts: FRB 180916}
\begin{abstract} The repeating FRBs 180916.J0158 and 121102 are visible
during periodically-occuring windows in time. We consider the constraints on 
internal magnetic fields and geometry 
if the cyclical behavior observed for FRB~180916.J0158 and FRB 121102 is due 
to precession of magnetars. In order to frustrate vortex line pinning we argue that
internal magnetic fields must be stronger than about $10^{16}$ Gauss, which is large
enough to prevent superconductivity in the core and destroy the crustal lattice structure.
We conjecture that the magnetic field inside precessing magnetars has three components,
(1) a dipole component with characteristic strength $\sim 10^{14}\,\Gauss$; (2) a toroidal 
component with characteristic strength $\sim 10^{15}-10^{16}\,\Gauss$ which only occupies
a modest fraction of the stellar volume; and (3) a disordered field with
characteristic strength $\sim 10^{16}\,\Gauss$. The disordered field is primarily responsible
for permitting precession,  which stops once this field component decays away, which we
conjecture happens after $\sim 1000$ years. Conceivably, as the disordered component
damps bursting activity diminishes and eventually ceases. We model the quadrupolar magnetic distortion 
of the star, which is due to its ordered components primarily, as triaxial and very likely
prolate. We address the question of whether or not the spin frequency ought to
be detectable for precessing, bursting magnetars by constructing a specific model in which
bursts happen randomly in time with random directions distributed in or between cones relative
to a single symmetry axis. Within the context of these specific models, we find that there 
are precession geometries for which detecting the spin frequency is very unlikely.
\end{abstract}

\newcommand{\lvhh}{\hat\ell}  

\newcommand{\muhatt}{\hat\mu} 
\section{Introduction}

The relatively long 16.4 day period of FRB~180916.J0158 \citep{2020Natur.582..351C}
and the even longer 160 day period of FRB 121102 \citep[e.g.][]{2021MNRAS.500..448C}
suggest precession of magnetars deformed by strong internal magnetic fields
\citep{2020ApJ...895L..30L, {2020ApJ...892L..15Z}}.
However, to date no evidence for a spin period has been reported for either of these FRBs \citep[e.g.][]{{2018ApJ...866..149Z},{2021arXiv210708205L}}. 
One possibility is that not enough bursts have been detected yet for either FRB to reveal their spin frequencies, presuming that the underlying engine
is a magnetar. 
%
But a second
possibility is that the physical nature of the repeating busts might prevent detecting a spin frequency even in upcoming surveys that detect
far larger numbers of individual outbursts. 

The important phenomenological questions motivating this paper are:
\begin{enumerate}
\item Should the spin period be detectable in FRBs that reappear periodically because of precession ?
\item Is it possible for there to be no evidence for either a spin period or a precession period for FRBs associated with a precessing magnetar ?
\end{enumerate}

Recently, evidence for a short $\approx 0.2$ second period has been presented by \cite{2021arXiv210708463T} from analysis of
the light curve of a {\sl single} outburst lasting $\approx 4$ seconds. This report lends urgency to addressing these two questions, and raises other
issues we shall not address here, such as whether or not the 0.2 second period is due to magnetar spin, and, if it is, what the implications are for
spindown, internal magnetic fields and precession.

In order to address these two questions, we first examine what the detection of precession tells us about the internal magnetic fields of the
magnetars presumed to be the sources of the FRBs.  \cite{1977ApJ...214..251S} showed that the pinning of (crustal) superfluid neutron vortex lines can
prevent slow precession, and \cite{2003PhRvL..91j1101L} showed that pinning of neutron vortices to flux tubes associated with proton superconductor
is likely wherever superfluid and superconductor coexist in the core of a neutron star. Moreover, for a neutron star rotating with period $P$ and
precessing with period $P_p$ the moment of inertia of the region where neutron vortices are pinned must be $\lesssim P/P_p\approx 10^{-7}P({\rm s})
/(P_p/100\,{\rm d})$ times the total moment of inertia of the star for slow precession to be possible. We can't rule out that FRB 121102 and
FRB~180916.J0158 are both fine-tuned to the accuracy necessary to permit slow precession. However, we regard it as far likelier that the magnetic
fields in the interiors of these magnetars are large enough to destroy proton superconductivity (and perhaps even neutron superfluidity).

\S \ref{sec:fields} is devoted to discussing constraints on the internal magnetic fields that would be consistent with precession.
We propose a specific model for the magnetic field that has three distinct components: in order of typical magnetic field strength these are a
dipole field, a toroidal field, both of which are ordered, and a disordered field. We develop this model in \S\ref{sec:magprec}, where we are
led inevitably to the conclusion that the quadrupole distortion of the star is triaxial, and most likely somewhat prolate. We also propose that a
magnetar may only precess for a relatively short portion of its life lasting perhaps 1000 years.

\S\ref{sec:triaxprec} develops results on triaxial precession necessary for the more phenomenological modelling done in \S\ref{apptofrbs}. In particular,
we show that rather large amplitude precession can be excited with relatively little fractional expenditure of magnetic energy, a natural consequence of the fact
that magnetic energy is sustantially larger than rotational energy in magnetars. We also consider two distinct types of effects due to spindown. In
\S\ref{sec:timingmodel} we develop the timing model relating observer time to precession phase when spindown is included. There are two effects, the
familiar long term spindown but also a cyclical effect specific to precessing pulsars that has period $P_p$ \citep{1993ASPC...36...43C}. 
In \S\ref{sec:long_term_evolution} we investigate the secular effect of spindown on the precession amplitude and phase, generalizing 
work done by \cite{1970ApJ...160L..11G} for oblate
axisymmetric precession to triaxial precession. We outline a simple phase diagram for this more complicated problem that is more complex than what arises
for oblate, axisymmetric precession.

Finally, in \S\ref{apptofrbs} we develop a very specific model in which we assume that FRBs are tied to magnetar outbursts that occur randomly in time
and point in random directions about some reference axis, which we take to be (but need not be) the magnetic dipole axis. We show that it is impossible
to detect either the spin frequency or the precession period if the outbursts can point in any direction, which is not a big surprise. 
However, we also find that the spin frequency ought to be easy to detect in some cases and much harder in others, depending on specific characteristics
of the precession model and the distribution of beam directions of the outbursts.

From a qualitative point of view, we offer two simple reasons that the outbursts underlying FRBs may occur randomly in time. Although tautological, 
one explanation is that the physical mechanism triggering the bursts simply is stochastic temporally, with burst directions that are random within some boundaries. 
Another is that the times between burst triggers are irregular but correlated, perhaps because there is 
a characteristic time for the burst phenomenon to reload, but associated with each outburst is a random
time offset, possibly as large as the spin period, related to where the burst is triggered within the magnetar magnetosphere. The bursts may
point in a large range of directions 
because they involve plasma moving relativistically along open magnetic field lines, 
leading to highly focussed energy output
in directions ranging from close to the magnetic dipole axis to perpendicular to the light cylinder. 
Alternatively, bursts may originate from a set of distinct, concentrated regions in the magnetosphere of a magnetar
that turn on and off stochastically, with each region beaming energy outward in a different direction.

In a companion paper we address the challenge of uncovering an underlying FRB spin frequency in a more general, phenomenological way that does not rely
on as specific a model for bursts from rotating magnetars as we develop in \S\ref{apptofrbs}. The model presented in this paper can be regarded as a definite physical
set up that realizes the general conditions for hiding the spin frequency of an FRB-inducing magnetar developed in the companion paper.
\newpage

\section{Internal Magnetic Fields and Triaxial Precession}
\label{sec:triax}

%

\subsection{Internal Magnetic Fields that Permit Slow Precession}
\label{sec:fields}

Previous work has focussed primarily on precession arising from oblate axisymmetric 
distortion due to magnetic stresses \citep[e.g.][]{{2020ApJ...895L..30L}, {2020ApJ...892L..15Z}}. 
Here, we examine what internal magnetic structure may be required for precession to occur, and highlight distinctive features 
that arise when the distortion is not axisymmetric and possibly prolate.

The internal magnetic structure of magnetars is not well-studied. In general, magnetohydrodynamic (MHD) studies of the magnetic fields in
normal conductors have shown that there are {\sl no} stable magnetic field configurations in barotropic normal fluids
\citep{{2012MNRAS.424..482L}, {2015MNRAS.447.1213M}}, but that stable, axisymmetric configurations may exist in stably stratified
fluids \citep{{2009A&A...499..557R}, {2013MNRAS.433.2445A}, {2015MNRAS.447.1213M}}. \cite{2009MNRAS.397..763B} and \cite{2013MNRAS.433.2445A}
argued that there may be stable magnetic field configurations in stably stratified stars whose poloidal fields are {\sl much weaker}
than their toroidal fields.
\cite{2016MNRAS.463.2542G} argued that the {\sl equilibrium} magnetic fields in non-barotropic normal fluid stars can be specified
freely if they are axisymmetric, but not if they are non-axisymmetric.  At the strong fields we envision, magnetization due to Landau quantization
of core electrons also affects stability \citep{{2021MNRAS.tmp.1692R},{2010ApJ...717..843S}}. Relativistic equilibria have been computed using realistic 
equations of state \citep[e.g.][]{{2001ApJ...554..322C},
{2008PhRvD..78d4045K},{2012MNRAS.427.3406F}}; equilibria were only found to exist if
the maximum internal magnetic field strength is $\lesssim 10^{18}$ G, which is a significant indication of limitations
imposed by overall dynamical stability, but does not assess MHD stability.  Of course, stability constraints are not necessarily relevant if 
the magnetic field is time-dependent, although presumably field configurations that are MHD unstable vary rather rapidly on timescales set by
the local Alfv\`en speed and the lengthscale of variation. 

Overall, these studies suggest that the internal magnetic fields of magnetars could be considerably stronger than 
their dipole (surface) magnetic field.

Another, rather different, argument also suggests strong internal magnetic fields. 
Once the core of the star cools below $\approx 10^9\degK$ core protons become
superconducting unless the internal magnetic field is stronger than the second critical field strength
\be
H_{{\rm c}2}=\frac{\Phi_0}{2\pi\xi_p^2}=\frac{(\pi m_p^\star\Delta_p/\pfp)^2c}{2e\hbar}
\approx\frac{9\times 10^{15}(\Delta_p/m_e)^2(m_p^\star/0.7m_p)^2}{(\pfp/\,100\MeV)^2}~\Gauss 
\label{Hctwo}
\ee
where $\Delta_p$ is the proton gap, $\pfp$ is the proton Fermi momentum, $m_p^\star$ is the proton effective mass, $\xi_p=\hbar\pfp/\pi m_p^\star\Delta_p$ is the
coherence length and $\Phi_0=\pi\hbar c/e$ is the flux quantum.
Proton gap calculations are complicated by many body effects at high densities
\citep{{2008PhRvC..78a5805Z}, {2014arXiv1406.6109G}, {2017PAN....80...77D},{2019NuPhA.986...18G}} but indicate that $\Delta_p\simeq 0.5\,\MeV\approx m_e$ near nuclear
density $\nnuc=0.16\,\fermi^{-3}$, where $\pfp\approx 100\,\MeV$; $\Delta_p$ decreases to zero at densities $\gtrsim 2\nnuc$. 
Of course, it is also possible that protons are superconducting
but magnetic field strengths at the inner boundary of the (normal) crust are below the first critical field strength
$$
H_{{\rm c}1}\approx \frac{2\times 10^{14}(\pfp/100\,\MeV)^3\ln\kappa}{(m_p^\star/0.7m_p)}~\Gauss,
\quad\quad\quad \kappa\approx\frac{4.8(m_p^\star/0.7m_p)^{3/2}(\Delta_p/m_e)}
{(\pfp/100\,\MeV)^{5/2}} ,
$$
in which case magnetic fields would not penetrate into the superconductor in equilibrium if entering from the outside. However, very likely
magnetic flux in the core is ``left over'' from before it cooled enough to become superconducting \citep[e.g.][]{1969Natur.224..673B}, 
in which case the proton superconductor in the core is in a ``mixed state'' with  $H_{{\rm c}2}>H>H_{{\rm c}1}$.  Quadrupolar deformations 
due to 
magnetic fields 
are 
\be
\epsmag=\frac{\beta_2HBR^4}{GM^2}\approx \frac{2\times 10^{-6}\beta_2B_{15}^2R_{10}^4}{M_{1.4}^2}\times\frac{H}{B}
\label{epsmagdef}
\ee
for magnetic induction $B=10^{15}B_{15}$ Gauss, stellar radius $R=10R_{10}$ km and mass $M=1.4M_{1.4}\msun$, where $H/B=1$ for a normal conductor but
$H/B>1$ for a type ii superconductor
\citep[e.g][]{{1975Ap&SS..33..215J},
{2002PhRvD..66h4025C},{2003MNRAS.341.1020W},{2013MNRAS.431.2986H},{2008MNRAS.383.1551A}}.
The parameter $\beta_2$ depends on the structure
of the neutron star and of its internal magnetic field, and represents how effectively the magnetic forces cause quadrupolar deformation.

The superconductor is type ii as long as $\kappa>1/\sqrt{2}$, which is the case throughout much of the region where protons are superconducting.
In a type ii superconductor, magnetic flux is organized into an array of thin flux tubes that have an areal density $eB/\pi\hbar c
\approx 5\times 10^{21}B_{15}\,\cm^{-2}$. 
If the neutrons are also superfluid
their vorticity would be confined into thin vortex lines with a much lower areal density $\mu_n\Omega/\pi\hbar\approx 3\times 10^4(\mu_n/m_n)/\Psec
\,\cm^{-2}$ for a rotation period $\Psec$ seconds, where $\mu_n/m_n$ is the neutron chemical potential in units of the neutron rest mass.
Unless the relative velocity between flux lines and vortex lines is big enough, which may be true
if the precession amplitude remains sufficiently large \citep{10.1046/j.1365-8711.2002.05726.x}, vortex lines will pin to flux lines 
\citep{2003PhRvL..91j1101L}, which frustrates slow precession \citep{1977ApJ...214..251S}. This problem can be avoided entirely if the core of the
neutron star is hot enough that neutrons remain normal: neutron gaps are $\sim 10-100\,\keV\simeq 10^{8-9}\degK$ 
\citep{{2008PhRvC..78a5805Z}, {2014arXiv1406.6109G}, {2017PAN....80...77D},{2019NuPhA.986...18G}}. Calculations by \cite{2018A&A...609A..74P}
indicate that the core of a magnetar may cool below ${\rm a\,few}\times 10^8{\rm K}$ within $\lesssim 100$ years of forming, so neutrons may be normal
 in some but not all of the core of a $\lesssim 100$ year old magnetar
unless the maximum neutron critical temperature in the core is lower than current estimates. 
But even a moderately small region where 
protons and neutrons are both superfluid could have an important impact on neutron star precession: for a precession period $\Ppdays$ days the moment of inertia $I_p$ of any region in the core where vortices are pinned
to proton flux tubes must be $\lesssim P/P_p\simeq 10^{-5}\Psec/\Ppdays$ times the total moment of inertia of the star \citep{1977ApJ...214..251S}.

{\it Here, we assume that the magnetic field in the core is strong enough to suppress superconductivity entirely.} This means that we suppose that 
 the total magnetic field strength $B=H>H_{{\rm c}2}$ throughout all of the core. For this to be true, the internal field strength must
be at least comparable to and probably larger than the dipole magnetic field at the stellar surface. This may be achieved if there are substantial 
non-dipolar components of the internal magnetic field, particularly toroidal components \citep[e.g.][]{2002PhRvD..66h4025C}.  \cite{2013MNRAS.433.2445A}
found stable, axisymmetric equilibria with toroidal fields $\lesssim 100$ times stronger than the poloidal fields. The deformation due to toroidal fields
can be axisymmetric, but if so is {\sl prolate} rather than {\sl oblate}.


Vortex line pinning may also be a problem in the neutron star crust, where neutron pairing is S-wave and superfluid gaps are larger, $\sim\MeV$
\citep{2014arXiv1406.6109G}; neutrons are likely to be superfluid down to densities close to neutron drip for temperatures $\sim{\rm a\,few}\times 10^8$K
\citep{2018A&A...609A..74P}.
Unpinning and repinning of neutron superfluid vortices to crustal nuclei have long been thought to be responsible
for the behavior of pulsar spins during and after rotational glitches \citep{{1975Natur.256...25A},{1982PMagA..45..227A},{1984ApJ...276..325A},{1993ApJ...403..285L}}.
Strong magnetic fields alter the equation of state because the degenerate electron gas becomes one dimensional once 
$$
\frac{p_F}{m_ec}\lesssim\sqrt{\frac{2eB\hbar}{m_e^2c^3}}=6.7\sqrt{B_{15}}
~;
$$
the equation of state of the inner crust is largely unaffected for $B\lesssim 10^{17}$ Gauss, although it stiffens considerably in the outer crust
\citep{2019PhRvC..99e5805M}. The crust ought to crystallize except at low densities for temperatures $\lesssim{\rm a\,few}\times 10^9\degK$
\citep{2020A&A...635A..84C}.
Neutron star precession can only persist in spite of potential pinning of neutron superfluid vortex lines to crustal nuclei or pasta phases 
\citep{{1983PhRvL..50.2066R},{1984PThPh..71..320H},{1993PhRvL..70..379L}} if the sustained
precession amplitude is large enough \citep{10.1046/j.1365-8711.2002.05726.x}. The complex topology of the nuclear pasta revealed by molecular
dynamics simulations \citep{2018PhRvC..98e5801S} could complicate pinning.

There are two other effects of superstrong crustal magnetic fields that should alter the physical conditions there, perhaps enabling precession to occur.
One effect is to shatter the crystalline crust, which can happen if $B^2/8\pi>\mu_{\rm el}$, where $\mu_{\rm el}$ is the elastic shear modulus.
Molecular dynamics simulations by \cite{2018PhRvL.121m2701C} indicate that the shear modulus of nuclear pasta is 
$\lesssim 10^{31}\erg\,\cm^{-3}\equiv(1.6\times 10^{16}\Gauss)^2/8\pi$, so crustal magnetic fields $\gtrsim 10^{16}$~\Gauss\ would shatter the crust. 
(See also \citealt{{1998PhLB..427....7P}}).

A second possibility presents itself for magnetic fields larger than the Clogston-Chandrasekhar limiting field strength
\citep{{1962PhRvL...9..266C},{1962ApPhL...1....7C}}
\be
B_{\rm CC}=\frac{\Delta_n}{\mu_n\sqrt{2}}\approx 1.2\times 10^{17}\Delta_n(\MeV) ~\Gauss
\label{BCC}
\ee
above which flipping the spin of one neutron can break a S-wave Cooper pair; here $\Delta_n$ is the neutron gap and $\mu_n$ is the neutron magnetic moment.
For such large magnetic field strengths, the uniform S-wave BCS superfluid condensate transitions to an inhomogeneous LOFF state \citep{{1974JETP...38..854L},
{1964PhRv..135..550F},{2018RPPh...81d6401K}}. Although the implications of such states in the crust have not been explored extensively, it is conceivable
that the inhomogeneous LOFF state behaves more like a crystal than a (super)fluid, which may permit precession to occur \citep{{2018qcs..confa1006L}}.
Moreover,  somewhat weaker magnetic fields may destroy the predominantly P-wave superfluidity of core neutrons for which
$\Delta_n\lesssim 100$ keV \citep[e.g.][]{2018ASSL..457..401H}.

In any event, we conclude that magnetic fields stronger than about $10^{16}\,\Gauss$ are necessary for slow precession. However, the precession period
is of order
\be
P_p\sim\frac{P}{\epsmag}\sim\frac{100\,{\rm d}\,P(\sec)}{10^7\epsmag}
\label{Pp}
\ee
which, in view of Eq. (\ref{epsmagdef}), suggests a quadrupolar deformation corresponding to $B\sim 10^{14}-10^{15}\,\Gauss$ for FRB 121102 and FRB 180916.J0158+65, 
which is too weak to prevent superconductivity of core protons according to Eq. (\ref{Hctwo}). We therefore propose that the magnetic fields inside these
magnetars consist of three components:
\begin{enumerate}
\item a dipole field with characteristic strength $B_D\sim 10^{14}\,\Gauss$;
\item a quadurpolar field with characteristic strength $B_T\sim (10^{15}-10^{16})\,\Gauss$ and a symmetry axis misaligned with the dipole moment;
\item a disordered magnetic field with characteristic strength $B_{\rm turb}\sim 10^{16}\,\Gauss$ strong enough to suppress superconductivity
but with large scale stresses that do not contribute significantly to the quadrupolar deformation of the star.
\end{enumerate}
The spindown timescale in this model is of order
\be
\tsd=\frac{I_0c^3P^2}{4\pi^2B_D^2R^6}\approx\frac{2\times 10^3\,{\rm y}\,\Iunit[P(\sec)]^2}{B_{D,14}^2R_{10}^6}
\ee
where $I_0=10^{45}\Iunit\,\gram\cm^2$ is the moment of inertia of the star and $B_D=10^{14}B_{D,14}\,\Gauss$; the expected ratio of spindown timescale
to precession period is
\be
\frac{\tsd}{P_p}=\left(\frac{c^3P}{4\pi^2GM}\right)\left(\frac{B}{B_D}\right)^2\left(\frac{I_0}{MR^2}\right)
\simeq\frac{7\times 10^4\,P(\sec)}{M_{1.4}}\left(\frac{B}{10B_D}\right)^2\left(\frac{I_0}{0.2MR^2}\right)~
\ee
so spindown is very slow compared with precession.  Below, we shall also suggest that $B_{\rm turb}$ decays via ambipolar diffusion
within $\sim 100-1000$ years. That would mean that if FRB 121102 and FRB 180916.J0158+65 have $P\simeq 10\,\sec$ they are both younger 
than their spindown ages and spinning close to their original rotational frequencies. However if they are spinning faster, with $P\simeq 1\,\sec$ then
they might be about as old as their spindown ages, have quadurpolar distortions $\epsmag\sim 10^{-7}$, and be of order halfway through their lifetimes 
as precessing neutron stars.

\subsection{Magnetic Precession}
\label{sec:magprec}

In this paper, we consider triaxial magnetic distortions that may be far from oblate. We shall see that such configurations lead to qualitatively 
new features for neutron star precession that may have distinctive observable consequences. 
Precession of a fluid star caused by magnetic distortions differs qualitatively from solid body precession even though mathematically the two are
the same. The inevitability of precession for stars with non-aligned spin and magnetic fields was originally pointed out by \cite{1958IAUS....6..169S},
and was studied extensively by Mestel and collaborators \citep{{1972MNRAS.156..419M},{1981MNRAS.195..979M},{1981MNRAS.196..491N}}. These studies also found that
there are slow, internal nonrigid motions in addition to 
uniform rotation which have been studied recently by \cite{2017MNRAS.467.4343L} for neutron 
stars with toroidal magnetic fields. Below, we neglect these motions, which are second order in small quantities although we recognize that they may
be significant for magnetic field evolution.

In a rotating, highly magnetic fluid the matter density is perturbed away from spherical symmetry. 
The moment of inertia tensor of the star only depends on the $l=2$ perturbations:
\be
I_{ij}=I_0\left[\delta_{ij}+\epsrot\left(\onethird\delta_{ij}-\Omhat_i\Omhat_j\right)+\epsmag m_{ij}\right]
\ee
where rotation is along the $\Omhat$ direction, $\epsrot$ and $\epsmag$ are the amplitudes of the $l=2$ distortions due to rotation and magnetic fields,
respectively, and $m_{ij}$ is symmetric and trace free (STF).
The stellar angular momentum is
\be
L_i=I_{ij}\Omega_j=I_0\left[\left(1-\frac{2\epsrot}{3}\right)\Omega_i+\epsmag m_{ij}\Omega_j\right]\equiv I_0'\left(\delta_{ij}+\epsmag'm_{ij}\right)\Omega_j~,
\label{LOmega}
\ee
where we use the summation convention. 
Since $\epsmag'=\epsmag[1+\Oscr(\epsrot)]$ we ignore the difference between $\epsmag$ and $\epsmag'$ below.
Invert Eq. (\ref{LOmega}) to get
$
\Omega_j=L_i(\delta_{ij}-\epsmag m_{ij})/I_0'
$
to first order in small quantities. Since $\epsmag m_{ij}$ is STF 
\be
\epsmag m_{ij}=-\frac{\eps}{2}\left(\delta_{ij}-\ehat_{3,i}\ehat_{3,j}\right)+\eps\ehat_{3,i}\ehat_{3,j}+\deps(\ehat_{1,i}\ehat_{1,j}-\ehat_{2,i}\ehat_{2,i})~;
\label{epsmij}
\ee
for an axisymmetric magnetic field $\deps=0$ but we regard this case as exceptional (although \cite{1958IAUS....6..169S} and \cite{1972MNRAS.156..419M}
and subsequent work focused on this situation). Since $\ehat_3$ is fixed in the rotating frame of reference
\be
\frac{d\ehat_3}{dt}
=
\left(1-\frac{\epsilon}{2}\right)
\frac{\Lvec\crossprod\ehat_3}{I_0'} - 
\frac{\deps(L_1\ehat_2 + L_2\ehat_1)}{I_0'}
\quad 
~\Rightarrow~\frac{d(\ehat_3\dotprod\Lvec)}{dt}=-\frac{2\deps L_1 L_2}{I_0'}
\ee
in the inertial frame. If the star is axisymmetric, precession is about the magnetic field axis of symmetry, as is well-known \citep[e.g.][]{{1958IAUS....6..169S},
{1970ApJ...160L..11G},{1972MNRAS.156..419M}}, but this is untrue for the more general non-axisymmetric case, where precession is more complicated.

For the intense magnetic fields we envision, the main cause of quadrupolar deformations are magnetic stresses. Even in non-barotropic stars, the 
magnetic field configurations that give rise to {\sl static} deformations are highly constrained \citep{2016MNRAS.463.2542G}. 
In such a star, 
the static structure is perturbed away from spherical symmetry by the Lorentz force density $\fvec_L$; to linear order
\be
0=-\grad\delta P+\rhat g_0(r)\delta\rho-\rho_0(r)\grad\dPsi+\fvec_L+\fvec_{NF}
\label{hydrobalance}
\ee
where $g_0(r)=-GM(r)/r^2$ is the gravitational acceleration in the unperturbed star, $\delta P(\rvec)$ is the pressure perturbation.
$\fvec_{NF}$ is due to non-fluid forces, and $\dPsi$ is the gravitational potential of the perturbation.
In the neutron star core, where $\fvec_{NF}=0$, axisymmetric static perturbations require that the toroidal field is
$$
\Bvec_T=B_T(r,\theta)\phihat=\frac{f(\psi)\phihat}{r\sin\theta}
$$
where $\psi(r,\theta)$ is the flux function of the poloidal field
$$
\Bvec_P=\frac{\grad\psi\crossprod\phihat}{r\sin\theta}~,
$$
but for nonaxisymmetric fields $f'(\psi)={\rm constant}$ \citep{2016MNRAS.463.2542G}.
Assuming that this
restriction holds, Eq. (\ref{hydrobalance}) is easy to solve in a non-barotropic star, where $\delta P$ and $\delta\rho$ are unrelated.
%

\cite{2013PhRvD..88j3005L} considered a specific example of a nonaxisymmetric field 
with dipole and toroidal fields that have different axes of symmetry $\muhat$ and $\that$, respectively, that lead to static deformations of the star. 
Below, we use a slightly different model for the dipole
and toroidal fields, and also include a disordered component. Assuming that the quadrupolar deformation due to the disordered component is relatively
small, 
%
magnetic forces due to the ordered dipole and toroidal fields 
result in a perturbed moment of inertia tensor 
\be
\delta I_{ij}=q_T\left(\frac{\delta_{ij}}{3}-{\that}_i{\that}_j\right)+q_D\left(\muhat_i\muhat_j-\frac{\delta_{ij}}{3}\right)~.
\ee
Both $q_T$ and $q_D$ are positive, and it follows
that the toroidal field promotes prolate deformations relative to a symmetry axis $\that$ whereas the dipole field promotes oblate deformations
relative to a symmetry axis $\muhat$. If we assume that 
\be
\muhat=\cos\beta\that+\sin\beta\uhat
\ee
in a right-handed $\uhat,\,\vhat,\, \that$ coordinate system then
we find that the eigenvalues of $\delta I_{ij}$ are
\be
\lambda_\pm=q_T\left[-\frac{1-d}{6}\pm\sqrt{\left(\frac{1-d}{2}\right)^2+d\sin^2\beta}\,\right]
~~~~~~~~~~\lambda_v=\frac{q_T(1-d)}{3}
\label{tilteddipoleeigs}
\ee
where we have defined $q_D=dq_T$. 
%

For the dipole field, we adopt a stream function
\be
\psi=B_Dh(r/R)r^2[1-(\muhat\dotprod\rhat)^2]
\ee
where the dimensionless function is
\be
h(x)=\frac{m(x)}{x^3}~
\ee
for a mass profile $M(r)=Mm(r/R)$. The magnetic field matches smoothly to an exterior vacuum dipole provided that both $\rho=0$ and $d\rho/dr=0$
at the stellar surface. Typically, the poloidal magnetic field vanishes somewhere along its equator, and is prone to
instability there.
In the Cowling approximation we find that
\be
q_D=\frac{2\mu^2}{GM}=\frac{2B_D^2R^6}{GM}~~~~~~~[{\rm Cowling}]
\label{qDC}
\ee
independent of the detailed density profile of the star, but including self-gravity changes $q_D$ by a factor $\sim 2$.

The toroidal field must vanish at the surface of the star in order to match to a vacuum exterior. (The toroidal fields in a pulsar magnetosphere are
much weaker than the internal toroidal fields we consider here.)  For $f'(\psi)={\rm constant}$ in the neutron star core, this important constraint can 
be satisfied in two ways that 
lead to very different values of $q_T$. 
\begin{enumerate}
\item The toroidal field may fill the core,
\be
\Bvec_T=\frac{B_Trh(r/R)\sin\theta\phihat}{R}=\frac{B_Tm(x)\sin\theta\phihat}{x^2}
\label{BTbig}
\ee
and plunge to zero in a thin boundary region where $\fvec_{NF}$ may be nonzero, thus loosening constraints on $\Bvec_T$, even if the crust is damaged 
severely by its strong magnetic field. 
The value of $q_T$ depends on the density profile; we adopt
\be
\rho(r)=\rho(0)\left(1-x^2\right)^2~\Rightarrow~m(x)=\frac{35x^3}{8}-\frac{21x^5}{4}+\frac{15x^7}{8}~{\rm and}~I_0=\frac{2MR^2}{9}
\label{densityprofile}
\ee
where $I_0$ is the moment of inertia of the spherical star; for this particular density profile including self gravity implies
\be
q_D\simeq\frac{3.19B_D^2R^6}{GM}~\Rightarrow~\epsmagD\simeq \frac{14.36B_D^2R^4}{GM^2}~.
\label{epsmagD}
\ee
To zeroth order in the shell thickness
\be
q_T=\frac{0.237B_T^2R^6}{GM}=\frac{0.188\btms R^6}{GM}~\Rightarrow~\epsmagT=\frac{0.845B_T^2R^4}{GM^2}
\label{qTfilled}
\ee
for this model, where $\btms=1.26B_T^2$ is the mean square toroidal field strength. The thin shell contributes about half of the deformation, which
may be unrealistic, so actual values could be as small as half as large.
\item The toroidal field may be confined to a limited valume if instead of Eq. (\ref{BTbig})
\be
\Bvec_T=B_T\phihat\left[\frac{m(x)\sin\theta}{x^2}-\frac{1}{x\sin\theta}\right]\Theta\left(\frac{m(x)\sin\theta}{x^2}-\frac{1}{x\sin\theta}\right)~,
\label{BTsmall}
\ee
which is the model used by \cite{2013PhRvD..88j3005L}. (See also \cite{2013MNRAS.433.2445A}, who introduced models of this type in their study of magnetic stability
in axisymmetry.) The field only occupies about 21\% of the stellar volume, and has a mean square $\btms=0.01B_T^2$ within 
this volume. In this case
\be
q_T=\frac{3.53\times 10^{-4}B_T^2R^6}{GM}=\frac{3.53\times 10^{-2}\btms R^6}{GM}~\Rightarrow~\epsmagT
=\frac{1.59\times 10^{-1}\btms R^4}{GM^2}~.
\label{qTtorus}
\ee
The quadrupole moment for this model is diminished severely because it is confined to such a small volume.
\item Interpreting these two models as extremes for quadrupolar distortion due to toroidal fields we estimate
\be
q_T\simeq \frac{(0.04-0.2)\btms R^6}{GM}
\label{qTest}
\ee
and therefore
\be
d=\frac{q_D}{q_T}\simeq \frac{(20-100)B_D^2}{\btms}=\frac{(0.2-1)(10B_D)^2}{\btms}~.
\label{dest}
\ee
Thus, $d\lesssim 1$ and may even be $\ll 1$; the quadrupolar distortions arising from ordered field are significantly triaxial, and very likely prolate.
\end{enumerate}

If the toroidal field occupies a small volume, as for Eq. (\ref{BTsmall}), then no matter how large $\btms$ is the field will
be incapable of suppressing proton superconductivity everywhere. Even if the field occupies much of the star, as in Eq. (\ref{BTbig}), it may not be strong 
enough to exceed $\hctwo$ even if it is much stronger than $B_D$. Moreover, as noted above $\Bvec_P$ is prone to instability in this model. Although
the toroidal field represented by Eq. (\ref{BTsmall}) can prevent the instability in axisymmetry if $\btms$ is large enough \citep{2013MNRAS.433.2445A}
we doubt that the tilted dipole model is stable ($\curl\Bvec_T\parallel\Bvec_P$ for both Eqs. (\ref{BTsmall}) and (\ref{BTbig}) in axisymmetry but
not in the tilted dipole model.) For these reasons we conclude that a precessing neutron star with internal fields that are stable on short timescales
ought to include a disordered component for with characteristic local field strength $\bturb>\sqrt{\btms}>B_D$. The disordered field may be a remnant
of the violent process that generated the strong internal magnetic fields \citep{{1993ApJ...408..194T},{2009MNRAS.397..763B}}.

The turbulent magnetic field is
\be
\Bvec_{\rm turb}=\grad N_1(\xvec)\crossprod\grad N_2(\xvec)
\label{Bturbmodel}
\ee
where the scalar functions $N_i(\xvec)$ are constant along field lines and are advected with the fluid in the limit of perfect conductivity. We can
think of $N_i$ as a pair of comoving field line labels. We assume that the turbulence is small scale locally but with a large scale bias, so that we can
expand
\be
N_i(\xvec)=\frac{1}{\sqrt{V}}\sum_{\kvec}S_i(\epsilon\xvec,k)\exp[i(\kvec\dotprod\xvec)+i\psi_i(\kvec)]
\label{Nimodel}
\ee
where $S_i(\eps\xvec,k)$ is a spectral function and $\epsilon$ is the ratio of the small scale that characterizes the local turbulence and the large
scale that characterizes the bias; $V$ is a normalization volume and  $\psi_i(\kvec)$ is a random phase. 
If we assume that $\psi_1(\kvec)=\psi_2(\kvec)$, which is plausible if the turbulent
field results from fluid motions that stretch, twist and fold individual field lines, then Eq. (\ref{Bturbmodel}) has a mean value $\sim\epsilon^2$ times
the characteristic local field amplitude, and there are magnetic forces $\sim\epsilon$ (corresponding to gradient of turbulent magnetic pressure), $\sim\epsilon^3$
and $\sim\epsilon^5$ (corresponding to the mean field). We assume that the local field is strong enough to destroy superconductivity, but that the forces
are too weak to have much effect on quadrupolar deformation. However, we do hope that the stresses can act as {\sl deus ex machina} to stabilize the ordered fields.


Magnetic fields in the core of a highly magnetic neutron star containing normal neutrons and protons decay via ambipolar diffusion:
\cite{1992ApJ...395..240R} estimate a decay timescale
\be
t_{\rm ambip}(L)\sim\frac{220\,{\rm y}\,(20Y)(T_8/3)^2[L(\km)]^2(n_b/\nnuc)^{2/3}}{B_{16}^2(L)}~,
\label{tambi}
\ee
for field varying on a length scale $L(\km)$ km,
where $n_b$ is baryon density, $\nnuc=0.16\,\fermi^{-3}$ is nuclear density, $Yn_b$ is the proton density, and $T=10^8T_8$.
(See also \cite{2011MNRAS.413.2021G}, \cite{2017MNRAS.465.3416P} and \cite{2017PhRvD..96j3012G}, Fig. 1.)
Neutron superfluidity would increase the decay time but as long as the core temperature is as high as
$\sim{\rm a\, few}\times 10^8$ K the normal neutron fraction will be considerable in much of the core, and Eq. (\ref{tambi}) remains true within a
factor of an order of magnitude or less. (Field decay heats the core, but given the steep $T$ dependence of neutrino cooling the core temperature is
not changed substantially.) The tangled component will decay on a timscale that depends on its fluctuation spectrum: if $B^2(L)\propto
L^\alpha$ the decay timescale $t_{\rm ambip}(L)\propto L^{2-\alpha}$, which implies faster decay on smaller scales for $\alpha<2$ and vice-versa;
theories of fully developed magnetic turbulence generally find $\alpha<1$ \citep{{1963AZh....40..742I},{1965PhFl....8.1385K},{1995ApJ...438..763G}}.
Plausibly, the tangled field decays away in a time $\lesssim
10^3$ y, after which substantial portions of the magnetar core become superconducting, which limits the time span during which a magnetar may precess
slowly. The larger scale ordered fields also decay as long as core protons are normal but since $L(\km)\simeq 10$ for these fields they may survive
relatively undiminished until protons become superconducting, after which ambipolar diffusion becomes ineffective. Ambipolar diffusion in the crust
is suppressed by neutron superfluidity. Magnetic field evolution in a magnetar crust involves an interplay among the Hall effect, Ohmic dissipation
and plastic flow \citep{2019MNRAS.486.4130L,{2016ApJ...833..189L}}, involving instabilities on timescales $\sim 10^3\,{\rm y}$ \citep{2020arXiv200103335G} 
and possibly evolution toward an attractor solution on timescales $\sim 10^5\,{\rm y}$ \citep{{2014MNRAS.438.1618G},{2014PhRvL.112q1101G}}. (The effect of
Landau quantization of crustal electrons on magnetar magnetic field evolution, which may be substantial, is being included for the first time in a forthcoming paper 
by \cite{{pbrinprep}}.)

\subsection{Triaxial Precession}
\label{sec:triaxprec}

Conservation of angular momentum is
\be
\frac{d\Lvec}{dt}=\frac{\dstar\Lvec}{dt}+\Omvec\crossprod\Lvec=\Nvec
\ee
where $\Lvec$ is the angular momentum and $\Nvec$ is the spindown torque; $\dstar/dt$ is time derivative in rotating frame. Substitute
$\Lvec=L\lvh$ where $L=\vert\Lvec\vert$ to get
\be
\frac{dL}{dt}=\frac{\dstar L}{dt}=\lvh\dotprod\Nvec\equiv N_\parallel~,
\label{spindowneqn}
\ee
and 
\be
\frac{\dstar\lvh}{dt}+\Omvec\crossprod\lvh=\frac{\Nvec-\lvh\lvh\dotprod\Nvec}{L}\equiv\frac{\Nvec_\perp}{L}~.
\label{eulerprecess}
\ee
For the spindown torque we adopt
\ba
\Nvec=-\frac{k\mu^2\Omega^2}{c^3}\left(\Omvec-a\Omvec\dotprod\muhat\muhat\right)~\Rightarrow~N_\parallel=-\frac{k\mu^2\Omega^2}{c^3}\left(\lvh\dotprod\Omvec
-a\Omvec\dotprod\muhat\lvh\dotprod\muhat\right)
\label{spindowntorque}
\ea
where $\muvec$ is the magnetic moment of the star, $k$ and $a$ are numerical constants $\sim 1$; for numerical evaluations we adopt $k=2$ and $a=1/2$,
which corresponds to a rate of energy loss $\Omvec\dotprod\Nvec=(\mu^2\Omega^4/c^3)(1+\sin^2\theta)$ where $cos\theta=\Omhat\dotprod\muhat$
\citep{2012ApJ...746...60L}. (The same spindown model was used in
\cite{2006MNRAS.365..653A}.) The angular velocity of rotation is
\be
\Omvec=L\left(\frac{\ehat_1\lvhh_1}{I_1}+\frac{\ehat_2\lvhh_2}{I_2}+\frac{\ehat_3\lvhh_3}{I_3}\right)
=\frac{L}{I_3}\left[\frac{I_3\ehat_1\lvhh_1}{I_1}+\frac{I_3\ehat_2\lvhh_2}{I_2}+\ehat_3\lvhh_3\right]
\ee
where $I_3>I_2>I_1$ are the moments of inertia along the principal axes of the quadrupolar distortion;
we define a parameter $0<e^2<\infty$, which measures the degree of triaxiality, in terms of which
\be
\frac{I_3}{I_2}=1+\frac{2\epsmag}{2+e^2},
~~~~~\frac{I_3}{I_1}=1+\frac{2(1+e^2)\epsmag}{2+e^2} ,
~~~~~e^2=\frac{I_3(I_2-I_1)}{I_1(I_3-I_2)}
\label{Is}
\ee
where $\epsmag$ is given by Eq. (\ref{epsmagdef}). {\sl Oblate} axisymmetric distortions ($I_1=I_2$) correspond to $e^2=0$; {\sl prolate} axisymmetric distortions
($I_3=I_2$) correspond to $e^2\to\infty$. For the tilted dipole model \citep[see Eq. (\ref{tilteddipoleeigs}) and][]{2013PhRvD..88j3005L}
\be
e^2=\frac{I_3(\vert 1-d\vert\sqrt{1+\Delta}+1-d)}{I_1(\vert 1-d\vert\sqrt{1+\Delta}-(1-d))}, ~~~~~\Delta=\frac{4d\sin^2\beta}{(1-d)^2}~.
\label{esqtd}
\ee
For $d\gg 1~\Rightarrow~\Delta\simeq 4\sin^2\beta/d\ll 1$ the dipole field dominates the quadrupolar distortion and 
Eq. (\ref{esqtd}) implies that $e^2\simeq\Delta/4\simeq \sin^2\beta/d\ll 1$. 
For $d\ll 1~\Rightarrow~\Delta\simeq 4d\sin^2\beta\ll 1$ the toroidal field dominates the quadrupolar distortion, and
Eq. (\ref{esqtd}) implies that $e^2\simeq 4/\Delta=1/d\sin^2\beta\gg 1$. For $\beta=0$, the axisymmetric case, $e^2=0$ if $d>1$ and $e^2=\infty$
for $d<1$.

The Euler equations have an exact conservation law
\be
\lvh_1^2+\lvh_2^2+\lvh_3^2=1
\label{lvhnorm}
\ee
because $\lvh$ is a unit vector, but there is also an approximate conservation law
\be
\Omvec\dotprod\lvh=\frac{\lvh_1^2}{I_1}+\frac{\lvh_2^2}{I_2}+\frac{\lvh_3^2}{I_3}\equiv\frac{2E_{\rm rot}}{L_0^2}~;
\label{secondconslaw}
\ee
$\dstar(\Omvec\dotprod\lvh)/dt=-\Omvec\dotprod\Nvec_\perp$ so $\Omvec\dotprod\lvh$ only varies appreciably on a timescale of order $1/\epsmag$ times the spindown time. In Eq. (\ref{secondconslaw}) the parameter $L_0$ is the magnitude of the stellar angular momentum at some reference start time, which could be the time when precession is excited.

The Euler equations allow steady state rotation about any of its three principal axes. By combining Eqs. (\ref{lvhnorm}) and (\ref{secondconslaw}) in three different ways appropriate to perturbation away from each principal axis we find
\ba
& &\frac{E_{\rm rot}}{L_0^2/2I_3}-1=\frac{2\epsmag[\lvh_1^2(1+e^2)
+\lvh_2^2]}{2+e^2}
\nonumber\\
& &\frac{E_{\rm rot}}{L_0^2/2I_2}-1=\frac{2\epsmag(\lvh_1^2e^2-\lvh_3^2)}{2+e^2+2\epsmag}
\nonumber\\
& &\frac{E_{\rm rot}}{L_0^2/2I_1}-1=
-\frac{2\epsmag[\lvh_2^2e^2+\lvh_3^2(1+e^2)]}{2+e^2+2\epsmag(1+e^2)}~.
\label{epertsforeachaxis}
\ea
Eq. (\ref{epertsforeachaxis}) shows that at a given angular momentum,
the lowest energy state is rotation about $\ehat_3$, the highest is
rotation about $\ehat_1$ and rotation about $\ehat_2$ is intermediate,
as is well-known. For axisymmetric oblate precession ($e^2=0$) the $\ehat_1$ and $\ehat_2$ directions are interchangeable, and precession about 
either one is unstable but stable about the symmetry axis $\ehat_3$, but for axisymmetric prolate precession ($e^2\to\infty$) 
the $\ehat_2$ and $\ehat_3$ directions are interchangeable, and precession is stable about either one and unstable about the symmetry axis $\ehat_1$.

Below, we will use the first of Eqs. (\ref{epertsforeachaxis}) to quantify the second conservation law by defining the energy perturbation above the minimum energy state to be
\be
E_{\rm rot}=\frac{L_0^2}{2I_3}\left[1+\frac{2\epsmag\Lambda^2(1+e^2)}
{2+e^2}\right]~\Rightarrow~\delta E_p=E_{\rm rot}-\frac{L_0^2}{2I_3}=\frac{L_0^2\Lambda^2\epsmag(1+e^2)}{I_3(2+e^2)}~,
\label{energychange}
\ee
where $\delta E_p$ is the extra energy associated with precession.
Using Eq. (\ref{energychange}) we write the conservation law as
\be
\Lambda^2(1+e^2)=\lvh_1^2(1+e^2)+\lvh_2^2~.
\label{conslawnewform}
\ee
Suppose precession is excited from its minimum energy state by injection of rotational energy $\delta E_p=\eta L_0^2/2I_3$. 
This is consistent with exciting precession with amplitude
\be
\Lambda^2(1+e^2)=\frac{(2+e^2)\eta}{2\epsmag}\equiv\frac{\eta}{\eta_{\rm crit}}~~~~~~\eta_{\rm crit}=\frac{2\epsmag}{2+e^2}~.
\ee
There are then two very different cases depending on how much energy is injected: if $\eta<\eta_{\rm crit}$ then $\Lambda^2(1+e^2)<1$ and
if $\eta>\eta_{\rm crit}$ then $\Lambda^2(1+e^2)>1$. We shall treat each of these cases, which have very different properties, separately.
Qualitatively, we shall see that $\Lambda^2(1+e^2)<1$ has well defined $e^2\to 0$ (axisymmetric, oblate) limiting dynamics whereas 
$\Lambda^2(1+e^2)>1$ has well defined $e^2\to\infty$ (axisymmetric, prolate) limiting dynamics.
Since $\epsmag\simeq\beta_2E_{\rm mag}/E_\star$
\be
\epsmag\simeq\frac{\beta_2E_{\rm mag}}{E_\star}
\label{eq:ratio_Emag_Estar}
\ee
where $\Emag\sim B^2R^3$ is the magnetic energy and $E_\star\sim GM^2/R$ is the binding energy of the neutron star, a more apt comparison is
\be
\frac{\delta E_{p}}{E_{\rm mag}}\simeq\frac{2\beta_2\epsrot(1+e^2)\Lambda^2}{2+e^2}
\label{eq:ratio_dErot_Emag}
\ee
where $\epsrot=L^2/2I_3E_\star$
is the rotational distortion of the star.

For a rotation period $\sim 1$ second we expect $\epsrot\sim I\Omega^2R/GM^2\sim 10^{-7}I_{45}R_{10}^3/M_{1.4}[P(\sec)]^2$ {\it so the energy required to excite
even high amplitude precession is only a small fraction of the magnetic energy of the star}. Even small changes in the magnetic field can engender
relatively large amplitude precession. To make this quantitative, suppose that a shearing event in the neutron star distorts the magnetic field
changing the moment of inertia of the star from $\Ivec$ to $\Ivec'=\Ivec+\DIvec$, where $\DIvec$ is STF. The eigenvalues of $\Ivec'$ are slightly
different than those of $\Ivec$, and its eigenvectors are rotated relative to the eigenvectors of $\Ivec$. If the eigenvalues and associated
eigenvectors of $\Ivec$ are $(I_i,\ehat_i)$, then to lowest order in $||\DIvec||$ the eigenvalues and eigenvectors of $\Ivec'$ are
\be
I_i'=I_i+\DI_{ii}~~~~~~\ehat_i'\simeq\ehat_i\left(1-\onehalf\sum_{j\neq i}\theta_{ij}^2\right)+\sum_{j\neq i}\theta_{ij}\ehat_j
~~~~~\theta_{ij}=\frac{\DI_{ij}}{I_i'-I_j'}~,
\label{Ichange}
\ee
normalizing the eigenvectors so that $\ehat_i'\dotprod\ehat_j'=\delta_{ij}$.
Since we expect $\DI_{ij}=s_{ij}\epsmag$, where $s_{ij}$ is STF with magnitude $\lesssim 1$, and $\vert I_i'-I_j'\vert\sim\epsmag$ the
rotations involve angles $\lesssim 1$, not $\sim\epsmag\ll 1$. Assuming that $\Lvec$ is conserved in the shearing event, its projection
along the rotated principal axes of $\Ivec'$ differs from its projection along $\Ivec$. For example, suppose that the star was rotating
without any precession at all along $\ehat_3$, the axis of largest moment of inertia, prior to the shearing event; then afterwards
\be
\lvh\simeq\ehat_1'\theta_{13}+\ehat_2'\theta_{23}+\ehat_3'\left[1-\onehalf\left(\theta_{13}^2+\theta_{23}^2\right)\right]~,\ee
and the star will precess. If angular momentum is conserved as the field rearranges itself, then the angular velocity changes during the
shearing event by
$$
\Delta\Omega_i=-\frac{\DI_{ij}\Omega_j}{I_i}
$$
working in the reference frame where $\Ivec$ is diagonal. The associated change in rotational energy is
\be
\Delta E_{\rm rot}=L_i\Delta\Omega_i=-\frac{L_i\DI_{ij}\Omega_j}{I_i}=-\Omega_i\DI_{ij}\Omega_j=-\epsmag\Omega_is_{ij}\Omega_j~.
\label{erotchange}
\ee
$\Delta E_{\rm rot}$ might be negative or positive, and is not equal to the extra energy in precession above the minimum energy state
corresponding to rotation about $\ehat_3'$, in part because the magnitude of the angular velocity changes as a result of the shearing event.
In rough order of magnitude $\vert\delta E_{\rm rot}\vert\lesssim\epsrot E_{\rm mag}$.

Phenomena associated with the spindown torque  include a cyclical variation over a precession cycle and a secular torque that develops 
very slowly compared with the precession period. We discuss these in Sections  \ref{sec:timingmodel} and 
\ref{sec:long_term_evolution} using a perturbative technique similar to \cite{1970ApJ...160L..11G} but generalized to triaxial precession. 
To zeroth order, we neglect spindown
effects, and Eq. (\ref{eulerprecess}) becomes
\be
 \frac{d\lvh_1}{d\phi}=-\frac{2\epsmag\lvh_2\lvh_3}{2+e^2}
~~~~
 \frac{d\lvh_2}{d\phi}=\frac{2\epsmag\lvh_1\lvh_3(1+e^2)}{2+e^2}
~~~~
\frac{d\lvh_3}{d\phi}=-\frac{2\epsmag\lvh_1\lvh_2 e^2}{2+e^2}
\label{eulerprecesscomps}
\ee
where $d\phi=(L/I_3)dt$ is differential spin phase.

From an observational standpoint, we are most interested in the 
motion of the direction from the star to the observer, $\nhat$, in the rotating frame of reference.
In the inertial frame, where $\lvh$ is independent of time to lowest order, let
\be
\nhat=\cosi\lvh+\sini\ehat_x~.
\label{nhatdef}
\ee
To project $\nhat$ into the rotating frame we use a standard Euler angle rotation \citep{1966qume.book.....G}:
(i) Rotate angle $\alpha\in[0,2\pi]$ about the 3 axis to get new axes $1',2',3'=3$,
(ii) Rotate angle $\beta\in[0,\pi]$ about the $2'$ axis to get new axes $1'',2''=2',3''=z$.
(iii) Rotate angle $\gamma\in[0,2\pi]$ about the $3''=z$ axis to get the axes $x,y,z$.
In terms of the angles $\alpha$, $\beta$ and $\gamma$ we get $\nhat=\nh_i\ehat_i$ where
\ba
& &\nh_1=\sini(\cos\alpha\cos\beta\cos\gamma-\sin\alpha\sin\gamma)-\cosi\cos\alpha\sin\beta
\nonumber\\
& &\nh_2=\sini(\sin\alpha\cos\beta\cos\gamma+\cos\alpha\sin\gamma)-\cosi\sin\alpha\sin\beta
\nonumber\\
& &\nh_3=\sini\cos\gamma\sin\beta+\cosi\cos\beta~.
\label{nhcomps}
\ea
If we define $\lvh=\zhat$ in the inertial frame we find that $\cos\beta=\lvh_3$ and
\be
\sin\alpha=-\frac{\lvh_2}{\sqrt{\lvh_1^2+\lvh_2^2}}~~~~~\cos\alpha=-\frac{\lvh_1}{\sqrt{\lvh_1^2+\lvh_2^2}}~.
\label{alphabeta}
\ee
Using $d\ehat_i/dt=\Omvec\crossprod\ehat_i$ for any of the principal axes we find that 
\be
\frac{d\gamma}{d\phi}=-1-\frac{2\epsmag[(1+e^2)\lvh_1^2+\lvh_2^2]}{(2+e^2)(\lvh_1^2+\lvh_2^2)}=-1-\frac{2\epsmag\Lambda^2(1+e^2)}{(2+e^2)(\lvh_1^2+\lvh_2^2)}
\label{dgammadphieqn}
\ee
using Eq. (\ref{conslawnewform}). 
%
%
Note that these results 
can be used for both $\Lambda\sqrt{1+e^2}<1$ and $\Lambda\sqrt{1+e^2}>1$.

In the rotating frame of reference,
$
\nhat=\cosi\lvh+\sini\left(\ehat_a\cos\gamma+\ehat_b\sin\gamma\right)
$
where $\ehat_a=(\lvh\crossprod\ehat_3)\crossprod\lvh/\sin\beta$ and $\ehat_b=\lvh\crossprod\ehat_3/\sin\beta$ are slowly varying unit vectors perpendicular to $\lvh$;
$\nhat$ rotates rapidly in the retrograde direction in the plane instantaneously perpendicular to $\lvh$.
For emission along a beam direction $\bhat$ the observed intensity depends on 
$$
\bhat\dotprod\nhat=\cosi\bhat\dotprod\lvh+\sini\left(\bhat\dotprod\ehat_a\cos\gamma+\bhat\dotprod\ehat_b\sin\gamma\right)~.
$$
Define $\bhat\dotprod\lvh=\cos\eta_b$, $\bhat\dotprod\ehat_a=\sin\eta_b\cos\psi_b$ and $\bhat\dotprod\ehat_b=\sin\eta_b\sin\psi_b$; then
\be
\bhat\dotprod\nhat=\cosi\cos\eta_b+\sini\sin\eta_b\cos(\gamma-\psi_b)~,
\label{bdotn}
\ee
where $\eta_b$ and $\psi_b$ vary during the precession cycle for a given $\bhat$. 

For ``pulsar-like'' behavior, beam directions are randomly distributed in a narrow cone around a dominant direction. Given unlimited
sensitivity, the observed intensity would be nearly periodic, with periodic timing residuals due
to precession $\lesssim 1$ radian of spin phase. The amplitude of the rapidly oscillating term in Eq. (\ref{bdotn}) is $\propto\sin\eta_b$, which varies during
the precession cycle. Presumably the observed intensity is a decreasing function of $\bhat\dotprod\nhat$, so the observed intensity has extrema when
\be
\frac{d\bhat\dotprod\nhat}{d\phi}=0~,
\label{bnzero}
\ee
which has solutions twice per cycle, only one of which corresponds to the maximum value of $\bhat\dotprod\nhat$.
Intrinsic intensity fluctuations and imperfect, time-varying and often unfavorable beaming due to precession 
turn out to render most pulses undetectable, but nevertheless the spin frequency would be discernible readily in this case.

For ``stochastic behavior'' in which outbursts occur randomly in time with a random distribution of beam directions it is much harder but not impossible
to uncover the pulse frequency. If the beams emit into narrow cones, Eq. (\ref{bdotn}) implies that most outbursts will not be seen but there will be a
bias favoring times when $\bhat\dotprod\nhat$ is near one. This bias imprints the effect of fast rotation on the times when outbursts happen, but only weakly,
so the spin frequency is only discernible after many bursts have been detected. We
develop a specific model for stochastic outbursts in \S \ref{apptofrbs}. An approximate analytic model that elucidates how information about the spin
frequency and precession period is imprinted on the modelled series of burst detection times may be found in Appendix \ref{app:analytic}.

Intermediate between these two extreme models would be one in which FRBs occur randomly in time but are triggered by exceptionally narrow beams within
a restricted range of possible directions. For an outburst occuring at a particular time, the associated FRB would only be seen if $\bhat$ is very
nearly parallel to $\nhat$, as determined from Eqs. (\ref{nhcomps}) with Eq. (\ref{alphabeta}), {\sl and} $\nhat$ is in the range of allowed beam
directions.

\begin{table}[h]
\caption{Precession Solutions}
\centering
\begin{tabular}{|c|c|c|}
\hline
	&      $0<\Lambda\sqrt{1+e^2}<1$ & $1<\Lambda\sqrt{1+e^2}<\sqrt{1+e^2}$\footnote{$\Lambda<1$ is required.}\\
\hline
\hline
	$q$ & $\frac{e\Lambda}{\sqrt{1-\Lambda^2}}<1$ & $\frac{e\Lambda}{\sqrt{1-\Lambda^2}}>1$\\
        $\Phi$\footnote{Precession phase. $F(\varphi|q)$ and $F(\varphi|1/q)$ are elliptic functions \citep[e.g.][ ]{1972hmfw.book.....A}} & $F(\varphi(\Phi)|q)=\int_0^{\varphi(\Phi)}\frac{d\varphi'}{\sqrt{1-q^2\sin^2\varphi'}}$ & $F(\varphi(\Phi)|1/q)=\int_0^{\varphi(\Phi)}\frac{d\varphi'}{\sqrt{1-\sin^2\varphi'/q^2}}$\\
        $\sn\Phi$ & $\sin[\varphi(\Phi)]$ & $\sin[\varphi(\Phi)]$\\
        $\cn\Phi$ & $\cos[\varphi(\Phi)]$ & $\cos[\varphi(\Phi)]$\\
        $\dn\Phi$ & $\sqrt{1-q^2\sn^2\Phi}$ & $\sqrt{1-\sn^2\Phi/q^2}$\\
        $\Phi_{p, {\rm cyc}}$\footnote{Precession phase per precession cycle.} &  $4F(\pi/2|q)$ & $4F(\pi/2|1/q)$\\
        $d\Phi/d\phi$ & $\frac{2\epsmag\sqrt{(1-\Lambda^2)(1+e^2)}}{2+e^2}=\frac{2\epsmag e\Lambda\sqrt{1+e^2}}{q(2+e^2)}$
        & $\frac{2\epsmag e\Lambda\sqrt{1+e^2}}{2+e^2}=\frac{2\epsmag q\sqrt{(1-\Lambda^2)(1+e^2)}}{2+e^2}$\\
        $\phi_{p, {\rm cyc}}$\footnote{Spin phase per precession cycle.} & $\frac{2(2+e^2)F(\pi/2|q)}{\epsmag\sqrt{(1-\Lambda^2)(1+e^2)}}=\frac{2q(2+e^2)F(\pi/2|q)}{\epsmag e\Lambda\sqrt{1+e^2}}$ &
        $\frac{2(2+e^2)F(\pi/2|1/q)}{\epsmag e\Lambda\sqrt{1+e^2}}=\frac{2(2+e^2)F(\pi/2|1/q)}{\epsmag q\sqrt{(1-\Lambda^2)(1+e^2)}}$\\
$\lvh_1$ & $\Lambda\cn\Phi$ & $\Lambda\dn\Phi$ \\
        $\lvh_2$ & $\Lambda\sqrt{1+e^2}\sn\Phi$ & $\frac{\sqrt{(1-\Lambda^2)(1+e^2)}\sn\Phi}{e}=\frac{\Lambda\sqrt{1+e^2}\sn\Phi}{q}$ \\
        $\lvh_3$ & $\sqrt{1-\Lambda^2}\dn\Phi$ & $\sqrt{1-\Lambda^2}\cn\Phi$\\
        $1+\frac{d\gamma}{d\phi}$ &$-\frac{\sqrt{1+e^2}d\Phi/d\phi}{\sqrt{(1-\Lambda^2)}(1+e^2\sn^2\Phi)}
        =-\frac{q\sqrt{1+e^2}d\Phi/d\phi}{e\Lambda(1+e^2\sn^2\phi)}$ &
        $-\frac{\Lambda\sqrt{1+e^2}d\Phi/d\phi}{e[\Lambda^2+(1-\Lambda^2)\sn^2\Phi]}=-\frac{\sqrt{1+e^2}d\Phi/d\phi}{e\Lambda(1+e^2\sn^2\Phi/q^2)}$\\
        &   & $=-\frac{\sqrt{1+e^2}d\Phi/d\phi}{q\sqrt{1-\Lambda^2}(1+e^2\sn^2\Phi/q^2)}$\\
	&   &      \\
\hline
\hline
	&    $e^2=0.0$ (Axisymmetric, Oblate) & $e^2=\infty$ (Axisymmetric, Prolate)\\
\hline
\hline
	$d\Phi/d\phi$ & $\epsmag\sqrt{1-\Lambda^2}$ & $2\epsmag\Lambda$ \\
        $\lvh_1$ & $\Lambda\cos\Phi$ & $\Lambda$ \\
	$\lvh_2$ & $\Lambda\sin\Phi$ & $\sqrt{1-\Lambda^2}\,\sin\Phi$\\
	$\lvh_3$ & $\sqrt{1-\Lambda^2}$ & $\sqrt{1-\Lambda^2}\,\cos\Phi$ \\
        $1+\frac{d\gamma}{d\phi}$ & $-\frac{d\Phi/d\phi}{\sqrt{1-\Lambda^2}}=-\epsmag$ & $-\frac{2\epsmag\Lambda^2}{\Lambda^2+(1-\Lambda^2)\sin^2\Phi}$\\
        &    & \\
\hline
\hline
\end{tabular}
        \label{table1}
\end{table}
Table \ref{table1} details the solutions of the Euler equations. Note that the solutions are continuous across the limiting case $\Lambda\sqrt{1+e^2}=1=q$,
but because the precession period diverges logarithmically as $q\to 1$ (from either side) the solutions are not really connected physically across $q=1$.

There are two different axisymmetric situations, $e^2=0$, which is oblate ($I_1=I_2<I_3$), and $e^2=\infty$, which is prolate ($I_3=I_2>I_1$); 
these solutions are listed in Table \ref{table1}. However, these are singular limiting cases: $q=e\Lambda/\sqrt{1-\Lambda^2}$ is identically zero
for $e=0$ and any value of $\Lambda\neq 1$ and is infinity for $e=\infty$ for any value of $\Lambda\neq 0$.

\begin{figure}[h]
\centering
\epsscale{1.0}
\plotone{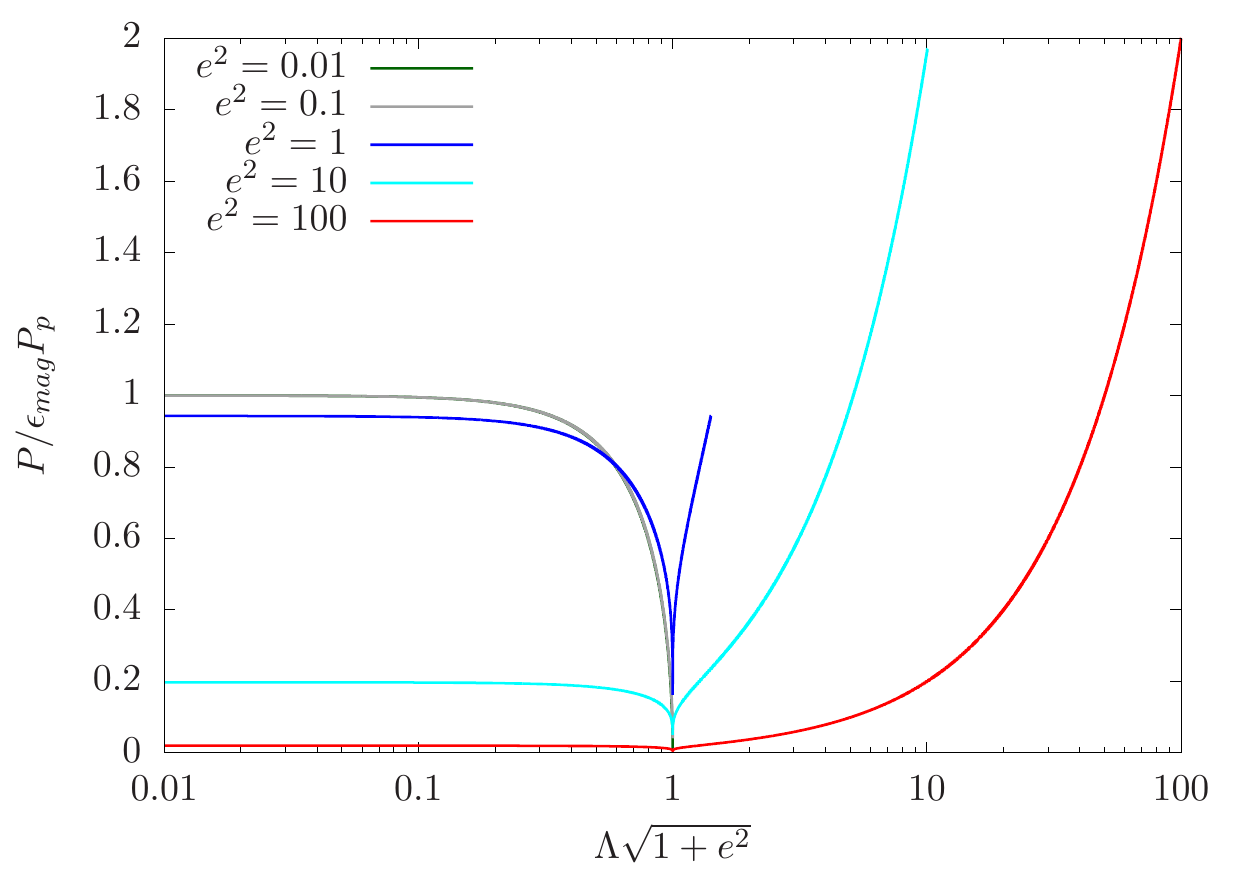}
\caption{$P/\epsmag P_p$ as a function of $\Lambda\sqrt{1+e^2}$ for $e^2=0.01,\,0.1,\,1,\,10$ and $100$.}
\label{fig:PvsPpplot}
\end{figure}

The ratio of the neutron star spin period $P$ to its precession period $P_p$ is 
\be
\frac{P}{P_p}=\frac{2\pi}{\phicyc}=\frac{\pi\epsmag e\Lambda\sqrt{1+e^2}}{2+e^2}\left\{\begin{array}{l} 1/qF(\pi/2|q)~~~~~\,[q<1] 
\\ 1/F(\pi/2|1/q)~~~~[q>1]\end{array}\right.
\label{PoverPp}
\ee
which is plotted in Fig. \ref{fig:PvsPpplot} for various values of $e^2$ as a function of $\Lambda\sqrt{1+e^2}$.
The smallest values of $P/P_p\epsmag$ are for $\Lambda\sqrt{1+e^2}<1$ and large $e^2$, except for the region right around $\Lambda\sqrt{1+e^2}=1$,
where $P/\epsmag P_p\to 0$ for all values of $e^2$. Since R1 and R3 both have very long $P_p$, Fig. \ref{fig:PvsPpplot} favors models with large
values of $e^2$, which implies that the toroidal component of magnetic field is significantly larger than the poloidal component, unless the
star is fortuitously close to $\Lambda\sqrt{1+e^2}=1$. 

\subsection{Periodic Timing Residuals from Precession Plus Spindown}
\label{sec:timingmodel}

Here, we derive the equation for $t(\phi)$, the functional dependence of time on spin phase, which we have defined previously as
$d\phi=(L/I_3)dt$; we use Eqs. \ref{spindowneqn} and \ref{spindowntorque} to {\sl zeroth order} in $\epsmag$ to get
\be
\frac{d\Omega}{d\phi}=-\frac{k\mu^2\Omega^2[1-a(\muhat\dotprod\lvh)^2]}{Ic^3}~\Rightarrow~\frac{d}{d\phi}\left(\frac{1}{\Omega}\right)
=\frac{d^2t}{d\phi^2}=\frac{k\mu^2[1-a(\muhat\dotprod\lvh)^2]}{Ic^3}~
\label{tofphieqn}
\ee
where $dt/d\phi=1/\Omega$ and $I_i\simeq I$ in this approximation. The solution to Eq. (\ref{tofphieqn}) is a continuous function $t(\phi)$ that 
exhibits the timing residuals due to spindown; for a single beam, which is appropriate for a precessing pulsar, we evaluate at $\{\phi_i\}$, the discrete set of 
spin phases where the pulses are beamed toward the observer optimally.

The dependence on precession phase arises from 
\be
[\muhat\dotprod\lvh(\Phi)]^2=\sum_i\muhat_i^2\lvh_i^2+2\sum_{i\neq j}\muhat_i\muhat_j\lvh_i\lvh_j
=\muhat_1^2\Lambda^2+\muhat_3^2(1-\Lambda^2)+\lvh_2^2\left(-\frac{\muhat_1^2+\muhat_3^2e^2}{1+e^2}+\muhat_2^2\right)
+2\sum_{i\neq j}\muhat_i\muhat_j\lvh_i\lvh_j~,
\ee
where we used the conservation laws, Eqs. (\ref{lvhnorm}) and (\ref{conslawnewform}), to separate out the constant term and isolate the 
dependence on $\lvh_2^2\propto\sn^2\Phi$. In evaluating $t(\phi)$ we must be careful to isolate secularly growing terms from terms that are
periodic over a precession cycle. We write the solution to Eq. (\ref{tofphieqn}) succinctly as
\ba
& &t(\phi)=t(0)+\frac{\phi}{\Omega(0)}\left[1-\frac{2ak\mu^2\Omega(0)}{Ic^3(d\Phi/d\phi)}\sum_{i<j}\muhat_i\muhat_jC_{ij}\right]
\nonumber\\& &
+\frac{k\mu^2\phi^2}{2Ic^3}\left\{1-a\left[\muhat_1^2\Lambda^2+\muhat_3^2(1-\Lambda^2)
+\langle\lvh_2^2\rangle\left(-\frac{\muhat_1^2+\muhat_3^2e^2}{1+e^2}+\muhat_2^2\right)\right]\right\}
\nonumber\\& &
-\frac{ak\mu^2}{Ic^3(d\Phi/d\phi)^2}\left[\left(\muhat_2^2-\frac{\muhat_1^2+\muhat_3^2e^2}{1+e^2}\right)P_{22}(\Phi)
+2\sum_{ij}\muhat_i\muhat_jP_{ij}(\Phi)\right]~.
\label{tofphiresult}
\ea
where $1/\Omega(0)=(dt/d\phi)_0$. Coefficents in Eq. (\ref{tofphiresult}) are given in Table \ref{table2}. The various averages
and functions in Table \ref{table2} are evaluated in Appendix \ref{app:funcs}. 

For calculations, it is convenient to express $t-t(0)$ in terms of the precession period $P_p$. Then $\phi/\Omega(0)P_p=\Phi/\Phicyc$
and the remaining terms all depend on the single nondimensional parameter
\be
\epsilon_{\rm sd}=\frac{\mu^2\Omega^2(0)P_p}{I_0c^3}=\frac{P_p}{\tsd}
\simeq\frac{2.3\times 10^{-4}B_{D,14}^2R_{10}^4(P_p/100\,{\rm d})}{[P(\sec)]^2M_{1.4}(I_0/0.2MR^2)}~
\label{epsddef}
\ee
because Eq. (\ref{tofphieqn}) may be written in the form
$$
\frac{d^2(t/P_p)}{d\Phi^2}=\frac{k\epsilon_{\rm sd}[1-a(\muhat\dotprod\lvh)^2]}{\Phicyc^2}~.
$$
From Eq. (\ref{tofphiresult}) we see that in addition to the apparent frequency
shift $\Oscr(\epsmag)$ arising from precession there is another apparent frequency shift $\Oscr(\epsilon_{\rm sd})$.
\begin{table}[h]
\caption{Coefficients in Timing Model}
\centering
\begin{tabular}{|c|c|c|}
\hline
 &  $q<1$ & $q>1$\\
\hline
\hline
        $\langle\lvh_2^2\rangle$ & $\Lambda^2(1+e^2)\langle\sn^2\Phi\rangle$ & $(1-\Lambda^2)(1+1/e^2)\langle\sn^2\Phi\rangle$\\
	$C_{12}$ & $\frac{\Lambda^2\sqrt{1+e^2}(1-\langle\dn\Phi\rangle)}{q^2} $ & $\Lambda\sqrt{(1-\Lambda^2)(1+1/e^2)}$\\
        $C_{13}$ &  $0$ & $0$\\
        $C_{23}$ & $\Lambda\sqrt{(1-\Lambda^2)(1+e^2)}$ & $q^2(1-\Lambda^2)\sqrt{1+1/e^2}(1-\langle\dn\Phi\rangle)$\\
        $P_{12}$ & $\Lambda^2\sqrt{1+e^2}C_2(\Phi|q)$ & $-\Lambda\sqrt{(1-\Lambda^2)(1+1/e^2)}C_4(\Phi|1/q)$\\
        $P_{13}$ & $\Lambda\sqrt{1-\Lambda^2}C_3(\Phi|q)$ & $\Lambda\sqrt{1-\Lambda^2}C_3(\Phi|1/q)$ \\
        $P_{23}$ & $-\Lambda\sqrt{(1-\Lambda^2)(1+e^2)}C_4(\Phi|q)$ & $(1-\Lambda^2)\sqrt{1+1/e^2}C_2(\Phi|1/q)$\\
        $P_{22}$ & $\Lambda^2(1+e^2)C_1(\Phi|q)$ & $(1-\Lambda^2)(1+1/e^2)C_1(\Phi|1/q)$\\
\hline
\end{tabular}
        \label{table2}
\end{table}
The amplitude of the cyclical terms is of order
\be
\Dt_{{\rm sd, cyc}}=\epsilon_{\rm sd}P_p=\frac{P_p^2}{t_{\rm sd}}
\ee
and the cyclical shift in pulse phase due to spindown is of order
\be
\Omega(0)\Dt_{{\rm sd, cyc}}=\frac{2\pi \epsilon_{\rm sd}P_p}{P}=\frac{2\pi P_p^2}{Pt_{\rm sd}}
\label{sdphaseosc}
\ee
which can be large for 
\be
\epsilon_{\rm sd}\gtrsim\epsilon_{{\rm sd},1}\equiv\frac{P}{P_p}\sim\epsmag
\label{epsdcrit}
\ee
\citep{1993ASPC...36...43C}. The secular terms $\propto\Phi^2$ also become progressively more important for $\epsilon_{\rm sd}>\epsilon_{{\rm sd}, 1}$
and, if large enough, may frustrate searches for the underlying spin period of the precessing magnetar in models based on stochastic outbursts. Eqs. (\ref{epsddef})
and (\ref{epsdcrit}) imply that
\be
\frac{\epsilon_{\rm sd}}{\epsilon_{{\rm sd},1}}=\frac{\mu^2\Omega^2(0)P_p^2}{I_0c^3P}
=\frac{1.9\times 10^3B_{D,14}^2R_{10}^4(P_p/100\,{\rm d})^2}{[P(\sec)]^3M_{1.4}(I_0/0.2MR^2)}~
\label{epssdoverepsdone}
\ee
which is between $\sim 5B_{D,14}^2$ and $\sim 5000B_{D,14}^2$ if $P_p=160\,{\rm d}$ (FRB 121102) and between 
$\sim 0.05B_{D,14}^2$ and $\sim 50B_{D,14}^2$ if $P_p=16.4\,{\rm d}$ (FRB 180916.J0158+65)
for $1/P\sim 0.1-1\,\sec^{-1}$.


\subsection{Secular Evolution of Precession via $\Nvec_\perp$}
\label{sec:long_term_evolution}

We now consider how precession evolves as a consequence of spindown, generalizing \cite{1970ApJ...160L..11G} to cases with
$e^2\neq 0$.  As in \cite{1970ApJ...160L..11G} we consider effects to lowest order in $\epsmag$.
We generalize the solutions to the Euler equations to include slow evolution of
the amplitude parameter $\Lambda=\Lambda(\epst)$, as was done by \cite{1970ApJ...160L..11G}, but also include a slowly varying
phase shift by replacing $\Phi=(d\Phi/d\phi)\phi$ with $\Phitil(t)=\Phi+\psi(\epst)$. This phase shift is required for triaxial precession
evolving via spindown.
Here $\epsilon=ak\mu^2(L/I_3)^2/c^3I_3$ is roughly the
inverse spindown time. 
We assume that the spindown time is long compared with the precession timescale, a necessary condition for a perturbative treatment;
this assumption fails at early times, or if $\Lambda\sqrt{1+e^2}\to 1$. 

We start by considering $\Lambda\sqrt{1+e^2}<1$, which is favored for long precession periods, and is the expected state if precession is excited
from rotation about $\ehat_3$ with relatively low $\delta E_p$.
Averaging over precession phase we find
\ba
& &\frac{1}{q}\frac{dq}{dt}
=\frac{ak\mu^2(L/I_3)^2}
{c^3I_3}\left\{\muhat_1^2[2-(1+q^2)\snsqav]+\muhat_2^2[1+(1-2q^2)\snsqav]-(1-q^2\snsqav)\right\}
\nonumber\\
& &\frac{d\psi}{dt}=\frac{ak\mu^2(L/I_3)^2\muhat_1\muhat_2}{c^3I_3\sqrt{1+e^2}}\left[\left\langle\frac{\cn^2\Phi}{\dn\Phi}\right\rangle
-(1+e^2)\left\langle\frac{\sn^2\Phi}{\dn\Phi}\right\rangle\right]~.
\label{secevprecess}
\ea
where we used
\be
\frac{dq}{q}=\frac{d\Lambda}{\Lambda}-\frac{d\sqrt{1-\Lambda^2}}{\sqrt{1-\Lambda^2}}=\frac{d\Lambda}{\Lambda}+\frac{\Lambda d\Lambda}{1-\Lambda^2}
=\frac{d\Lambda}{\Lambda(1-\Lambda^2)}-\frac{d\sqrt{1-\Lambda^2}}{\Lambda^2\sqrt{1-\Lambda^2}}~.
\label{dqoverq}
\ee
Eq. (\ref{secevprecess}) reduces to the results in Goldreich (1970) for $e^2=0$, the axisymmetric oblate case, for which $\cnsqav=\snsqav=\onehalf$ and
$\dn\Phi=1$, after replacing $dq/q\to d\Lambda/\Lambda(1-\Lambda^2)$ using Eq. (\ref{dqoverq}); for that case, there is no phase shift $\psi$.
The stability condition implied by the first
of Eqs. (\ref{secevprecess}) is more complicated than what was found by Goldreich (1970) for the axisymmetric, oblate case: 
there is a separatrix that is an ellipse in the $\muhat_1-\muhat_2$ plane whose
axes depend on $q$, so that, for given values of $e^2$ and $\muhat_1^2$ and $\muhat_2^2$, there is a fixed point at a specific value of $q$.

\begin{figure}[h]
\centering
\epsscale{1.0}
\plotone{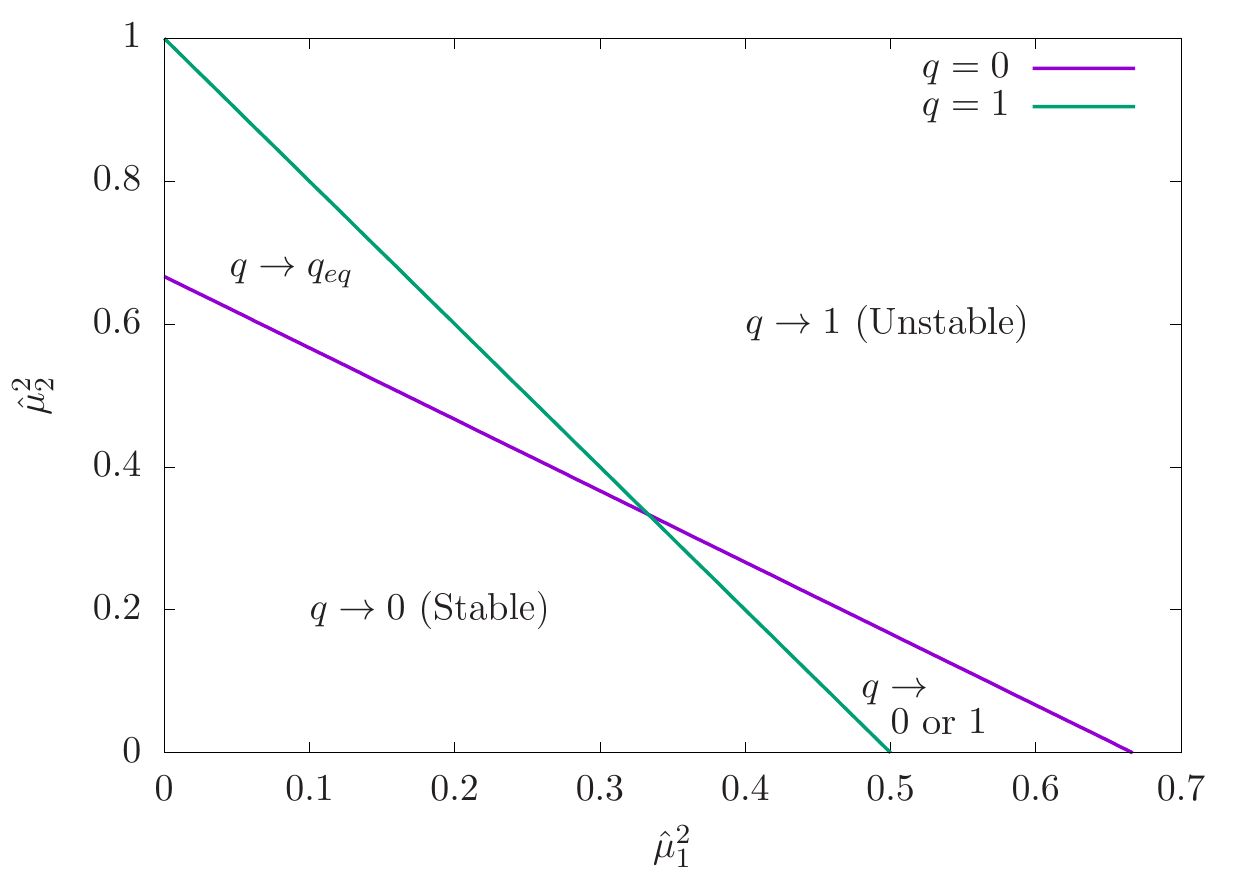}
\caption{Stability boundaries for $q=0$ and $q=1$ and outcomes for secular evolution of $q$. 
}
\label{fig:stabilityplot}
\end{figure}

By contrast, for the axisymmetric oblate case, the ellipse degenerates into a circle
\be
\muhat_1^2+\muhat_2^2=\frac{2}{3}
\label{oblatecircle}
\ee
irrespective of the value of $\Lambda$. In this case, $\Lambda$ grows as long as the magnetic moment configuration of the star is outside this circle.
The growth halts once $\Lambda\to 1$, where $d\Lambda/dt$ also goes to zero.
Inside the circle, $\Lambda$ decreases, reaching $\Lambda=0$ asymptotically.

For the triaxial case $q$ cannot grow beyond $q=1$; since $\snsqav=1$ for $q=1$, Eq. (\ref{secevprecess}) implies $dq/dt=0$. The stability curve (infinitesimally)
close to $q^2=1$ is 
\be
2\muhat_1^2+\muhat_2^2=1~.
\label{stabilityqone}
\ee
The two curves intersect at $\muhat_1^2=\muhat_2^2=1/3$; in fact all of the stability boundaries intersect at this point
since $\cnsqav=1-\snsqav$ and $\dnsqav=1-q^2\snsqav$. These two bounding stability lines are depicted in Fig. \ref{fig:stabilityplot}. 
The stability boundaries for all other values of $q$ are between these two lines, pivoting about their intersection point at $\muhat_1^2=\muhat_2^2=1/3$.

Fig. \ref{fig:stabilityplot} depicts evolution outcomes for various regions in the $\mu_1^2-\mu_2^2$ phase plane. The region marked ``$q\to 1$'' is unconditionally
unstable: if precession is excited for $(\mu_1^2,\mu_2^2)$ in this region, $q$ grows on the spindown timescale until $q=1$, where growth ceases.
Conversely, the region marked ``$q\to 0$'' is unconditionally stable: if precession is excited for $(\mu_1^2,\mu_2^2)$ in this region, $q$ shrinks 
toward zero on the spindown timescale. In the two triangular regions between the $q=0$ and $q=1$ bounding curves $dq/dt=0$ at $q=q_{\rm eq}(\mu_1^2,\mu_2^2)$
for each $(\mu_1^2,\mu_2^2)$. In the region marked ``$q\to q_{\rm eq}$,'' $dq/dt<0$ for $q>q_{\rm eq}(\mu_1^2,\mu_2^2)$ and $dq/dt>0$ for 
$q<q_{\rm eq}(\mu_1^2,\mu_2^2)$, so if precession is excited in this region $q\to q_{\rm eq}(\mu_1^2,\mu_2^2)$ asymptotically as a result of spindown.
In the region marked ``$q\to 0$ or $1$'', $dq/dt>0$ for $q>q_{\rm eq}(\mu_1^2,\mu_2^2)$ so $q\to 1$ asymptotically if precession is excited with 
$q>q_{\rm eq}(\mu_1^2,\mu_2^2)$ and $dq/dt<0$ for $q<q_{\rm eq}(\mu_1^2,\mu_2^2)$ so $q\to 0$ asymptotically if precession is excited with 
$q<q_{\rm eq}(\mu_1^2,\mu_2^2)$. 

For precession in the $\Lambda\sqrt{1+e^2}>1$ regime we get
\ba
& &-\frac{1}{q}\frac{dq}{dt}
=\frac{ak\mu^2(L/I_3)^2}
{c^3I_3}\left\{\muhat_3^2\left[2-\left(1+\frac{1}{q^2}\right)\snsqav\right]
+\muhat_2^2\left[1+\left(1-\frac{2}{q^2}\right)\snsqav\right]-\left(1-\frac{\snsqav}{q^2}\right)\right\}
\nonumber\\
& &\frac{d\psi}{dt}=\frac{ak\mu^2(L/I_3)^2\muhat_3\muhat_2e}{c^3I_3\sqrt{1+e^2}}\left[\left\langle\frac{\cn^2\Phi}{\dn\Phi}\right\rangle
-\left(1+\frac{1}{e^2}\right)\left\langle\frac{\sn^2\Phi}{\dn\Phi}\right\rangle\right]~,
\label{secevprecesshi}
\ea
where we used Eq. (\ref{dqoverq}) again.  Eq. (\ref{secevprecesshi}) may be obtained from Eq. (\ref{secevprecess}) with the substitutions
$\muhat_1^2\to\muhat_3^2$, $e^2\to 1/e^2$ and and $q\to 1/q$. The evolution scenarios for $1/q$ analogous to those for $q$ shown in Fig. \ref{fig:stabilityplot} 
may be derived using this mapping.

\section{Application to Fast Radio Bursts}
\label{apptofrbs}

\subsection{A Random Burst Model}
\label{randomburstmodel}

To this point, we have focussed on the combined effects of precession and spindown on observations of emission along a single beam in which emitted intensity
is determined entirely by $\bhat\dotprod\nhat$. For application to FRBs we develop a different model in which multiple beams pointing in random directions
fire at random times with random intrinsic intensities.


To address these questions we simulate an ideal observing program consisting of daily observations over a total observing time
lasting $n_{\rm p,cycle}$ precession cycles. In our idealized observing campaign, each daily observation starts one day after the beginning 
of the previous one and lasts $f_{\rm obs}$ days (2.4 hours).
We assume that bursts occur at a uniform rate throughout the duration of the observing program and that $n_{\rm bursts}$ occur
during the total time $f_{\rm obs}n_{\rm p,cycle}P_p$ of the observations. We input $P_p$ in days, so that the number of observing
days is the integer part of $P_p$ plus one.

We also input the parameters of the precession model, $(\epsmag,\Lambda,e^2)$ from which we can compute the spin frequency $\nu_\star$.
We choose $\gamma(0)$ randomly in the range $[0,2\pi]$. 

The simulation starts by choosing the set $\{\Phi_j\}$ of individual burst precession phases; in the absence of spindown the corresponding
burst times are $t_j=P_p\Phi_j/\Phipcyc$, and even with spindown included the burst times only differ from these times by
$\Oscr(\epsilon_{\rm sd})$. For each simulation there are $n_{\rm bursts}$ bursts. Ultimately, only a small fraction of these are detectable.

For each $\Phi_j$ we next determine a beam direction $\bhat_j$ in the rotating frame of reference. We do this relative to a reference
beam whose direction we input. In the calculations presented here we assume that this reference direction coincides with the direction
of the dipole moment appearing in the spindown formula, $\muhat$. We assume that $\bhat_j$ is anywhere between two cones about $\muhat$ 
defined by $\cos\theta_{\rm min}\leq\bhat_j\dotprod\muhat\leq\cos\theta_{\rm max}$, adopting a uniform distribution for $\bhat_j\dotprod\muhat$
over this range and a uniform direction of azimuthal angles in $[0,2\pi]$. We could, of course, choose a different reference direction or
multiple reference directions among which bursts may switch. As should already be apparent, there are many parameters in this model, and
choosing a single reference direction simplifies the calculation somewhat. Our model does allow the reference direction to switch to a different one
with a probability $p_{\rm flip}=1-f_{\rm no\,flip}$, but the results reported here are for $p_{\rm flip}=0$.

The next step is to compute $\bhat_j\dotprod\nhat$ for each outburst. We do this by computing $\nhat(\Phi_j)$ in the rotating frame of
reference from Eq. (\ref{bdotn}). This requires choosing a value of the inclination angle $i$ defined in Eq. (\ref{nhatdef}), which we input.

Once we have the value of $\bhat_j\dotprod\nhat(\Phi_j)\equiv\bhat_j\dotprod\nhat_j$ for a given outburst we can decide whether or not that outburst is detectable.
As a first cut,
we discard all bursts for which $\bhat_j\dotprod\nhat_j<0$ since these point away from the observer. Because we assume that 
each beam has a FWHM $\theta_{\rm FWHM}$ we may be discarding some bursts that could be detectable, in principle, but as long as $\theta_{\rm FWHM}$
is not too large we expect that this is not an important source of inaccuracy in our conclusions. We do not discard beams that would be eclipsed by
the neutron star. To do that we would need to specify the radius $r_b$ from which emission originates for beam $b$; eclipses could occur for $\cos\theta_b<0$
and $\pi-\theta_b\lesssim R/r_b$. In most of our simulations we restrtict $\cos\theta_b\geq 0$.

We assume a Gaussian emission pattern for each beam with an intrinsic intensity
\be
I_j(\bhat\dotprod\nhat)=I_j(1)\exp[\kappa(\bhat_j\dotprod\nhat_j-1)]
\label{Ijmodel}
\ee
where, if the FWHM of the beam is $\theta_{\rm FWHM}$,
\be
\kappa=\frac{\ln 2}{1-\cos(\onehalf\theta_{\rm FWHM})}~;
\label{kappadef}
\ee
$I_j(1)$ is the peak intensity for outburst $j$ and 
\be
\int_0^1 d\mu I_j(\mu)=\frac{2\pi I_j(1)[1-\exp(-\kappa)]}{\kappa}
\ee
is the total intensity of the beam integrated over directions.
Eq. (\ref{Ijmodel}) would be the final answer
if all outbursts were equally intense intrinsically, but in general we expect a distribution of $I_j(1)$. To model this, we input a range of intrinsic
intensities, and assume that the distribution of intrinsic intensities $I_j(1)$ is uniform in $\ln I_j(1)$ over the corresponding logarithmic range 
with a mean value of one. (In this model we could choose a different mean value, but this would just introduce a multiplicative factor in each value of $I_j(1)$.) 
After selecting $\ln I_j(1)$ at random from this distribution we evaluate $I_j$ using Eq. (\ref{Ijmodel}). 

Given $\{I_j\}$ we can find the maximum value $I_{\rm max}$. We assume that only bursts greater than $I_{\rm min}=I_{\rm max}\times (I_{\rm min}/I_{\rm max})$
are detectable, where $I_{\rm min}/I_{\rm max}$ is another input parameter. We then have the sets $\{\Phi_j\}$ and $\{I_j/I_{\rm max}\}$ for the bursts.
The latter can immediately be turned into a plot of number of detected bursts per (daily) observing session versus precession phase, which provides a
simple visual indication of whether the data reveal or conceal the precession period.
The same data can be plotted as a cumulative distribution of observed burst intensities which we shall see is different from the inputted distribution
of intrinsic burst intensities.


The final step in our calculations is to 
determine conditions under which the spin frequency ought to be detected.
We do this by computing
\be
{\hat D}(\nu_\star)=\sum_j w_j\exp[2\pi i\nu_\star t(\Phi_j)]
\label{Ddef}
\ee
where $\nu_\star=\Omega(0)/2\pi$ is the (initial) spin frequency of the star and $t(\Phi_j)$ is computed from 
Eq. (\ref{tofphiresult}) for selected values of $\epsilon_{\rm sd}$. In Eq. (\ref{Ddef}) $\{w_j\}$ is a set of weights assigned to each detected burst.
To assess the evidence for a given $\nu$ we compute $\vert D(\nu)\vert$. For totally uncorrelated $t_j$
\be
\left\langle\vert\hat D(\nu_\star)\vert^2\right\rangle_{\rm uncorrelated}=\sum_j w_j^2
\ee
so we normalize the computed values:
\be
\vert\hat D(\nu_\star)\vert_{\rm normalized}\equiv\frac{\vert\hat D(\nustar)\vert}{\sqrt{\left\langle\vert\hat D(\nustar)\vert^2\right\rangle_{\rm uncorrelated}}}~.
\ee
In our calculations we weight each term in Eq. (\ref{Ddef}) equally, so that $w_j=1/N_{\rm bursts}$ for $N_{\rm bursts}$ detected bursts; with this
choice $\langle\vert D(\nu)\vert^2\rangle_{\rm uncorrelated}=1/N_{\rm bursts}$, and
\be
\vert\hat D(\nu_\star)\vert_{\rm normalized}\equiv\vert{\hat D(\nu_\star)}\vert\sqrt{N_{\rm bursts}}~.
\label{DofN}
\ee
Another plausible choice for $w_j$ would be $I_j/I_{\rm max}$. 
%
If the burst times are precisely periodic then $\vert\Dhat(\nustar)\vert=1$ and $\vert\Dhat(\nustar)\vert_{\rm normalized}=\sqrt{N_{\rm bursts}}$.
This remains true for $\nustar'=\nustar+\Delta\nustar$; the frequency shift associated with spindown, which is included in our calculation, is undetectable.  
For a single beam, there would be a systematic frequency shift $\Oscr(\epsmag)$ that depends on beam direction, but for multiple beams there is no systematic
shift.
The value of $\vert\Dhat(\nustar)\vert$ is unaffected by shifting the burst times
by a common time offset.
If burst times are random, the asymptotic probability distribution of $r=\vert\Dhat(\nustar)\vert\sqrt{N}$ is
\be
\frac{dp(r)}{dr}=2r\exp(-r^2)
\label{rayleigh}
\ee
independent of $N$. The mode of Eq. (\ref{rayleigh}) is $r=1/\sqrt{2}$ and the mean is $\sqrt{\pi}/2$.

In our models, we evaluate $D_d(\nustar)=\vert\Dhat(\nustar)\vert\sqrt{N_d}$ for each of $\{d\}$ days of observations during which $\{N_d\}$ bursts are
detected.  According to Eq. (\ref{rayleigh}) the probability that $D_d>r$ is $\exp(-r^2)$ if the bursts occur randomly. If observations are done on $M$
days the expected number of values of $D_d$ that exceed $r$ is $n(>r)=M\exp(-r^2)$, and the value of $r$ for which $(n>r)=1$ is
\be
r_1(M)=\sqrt{\ln M}~.
\label{ronedef}
\ee
The probability that no values of $\vert\hat D_d(\nustar)\vert\sqrt{N_d}>r_0$ are found at random is 
$$
p(r_0,M)=[1-\exp(-r_0^2)]^M
$$
so for a chosen value $p=p(r_0|M)$
\be
r_0(p,M)=\frac{1}{\sqrt{\ln(1-p^{1/M})}}=\frac{1}{\sqrt{\ln[1-\exp(\ln p)/M]}}\approx\frac{1}{\sqrt{-\ln p/M}}=\frac{r_1(M)}{\sqrt{(-\ln p)}}~;
\label{rzerodef}
\ee
$r_1(M)\approx r_0(1/e,M)$. Below we use $r_1(M)$ to assess the dectability of $\nustar$ over $M$ days by keeping track of the number of days for which
$\Dhat_d(\nustar)$ exceeds $r_1(M)$. 

Of course the observer will not know $\nustar$ in advance but we presume that he/she analyzes the data for a broad range of possible
spin frequencies including test values near $\nustar$. In our simulations, we compute $\{\Dhat_d(\nustar)\}$ for each of $M=512$ consecutive days, 
so $r_1(M)=\sqrt{\ln 512}=2.498$.
We focus on the day with the
largest value $\Dhat(\nustar)\sqrt{N_d}$ and for that day we calculate $\Dhat(\nu)\sqrt{N_d}$ for $N_{\rm freq}$ different frequencies. 
For small enough spacing between the test frequencies 
$\nustar$ ought to be very near one of the sampled frequencies; a value above $r_1(N_{\rm freq})$ is considered to be signficant. In the simulations
reported in Table \ref{tab:BM} we sample frequencies spaced by $\Delta\nu/\nu=10^{-5}$ Hz between $0.05$ Hz and $5$ Hz, a total of $N_{\rm freq}=460518$ 
frequencies, so $r_1(N_{\rm freq})=\sqrt{\ln 460518}=3.611$. Although we have only done frequency searches on the most promising 
day for each burst model, the spin frequency ought to be detectable on any day for which $\Dhat_d(\nustar)\sqrt{N_d}>r_1(N_{\rm freq})$, so 
we tabulate the number of such days.

In Appendix \ref{app:analytic} we develop an analytic theory of the probability of burst detections at a given time in our model. Eq. (\ref{Pdetect})
makes it clear that the probability depends on spin and precession frequency via $\nhat(\Phi)\dotprod\muhat$. Moreover, there is no time dependence at all
if the distribution of beam directions is isotropic. Thus, the observation of regular precession cycles by itself argues for a restricted range of beam directions.

\begin{table}[h]
\caption{Simulated Burst Models\footnote{Consecutive daily observations lasting 0.1 d each for 512 d, observer at $i=52^\circ$. A total of 1024000 outbursts. Beam width $\theta_{\rm FWHM}=20^\circ$. Intrinsic intensity range a factor of 1000; ratio of minimum to maximum observed intensities $I_{\rm min}/I_{\rm max}=0.01$. $\epsilon_{\rm sd}=0$.}}
\centering
\begin{tabular}{|c|c|c|c|c|c|c|c|c|c|}
\hline
	$P_p$(d) & $\epsmag$ & $\Lambda$ & $\nustar$ (Hz) & $\cos\theta_b$\footnote{Range of beam offsets axisymmetric relative to symmetry axis at $\ehat_3\cos\theta_\mu+\sin\theta_\mu(\ehat_1\cos\varphi_\mu+\ehat_2\sin\varphi_\mu)$ with $(\theta_\mu,\varphi_\mu)=(30^\circ,40^\circ)$.} & $D_{max}$(d)\footnote{Maximum value of $\hat D_d\sqrt{N_d}$ and day on which it occurs.} & $N_{\rm bursts}$(\%)\footnote{Total number of detectable bursts and fraction of total number of outbursts.} & $N_d(512)$\footnote{Number of days for which ${\hat D}_d\sqrt{N_d}$ exceeds $\sqrt{\ln 512}=2.498\ldots$.} & $N_d(N_{\rm freq})$\footnote{Number of days for which ${\hat D}_d\sqrt{N_d}$ exceeds $\sqrt{\ln N_{\rm freq}}=3.611\ldots$.} & Description \\
\hline
\hline
	160 & $10^{-6}$ & 0.2 & 0.1521 & [0.99,1] & 13.6 (411) &  37596 (3.7\%) & 399 & 384 & pulsar-like\\
	160 & $10^{-7}$ & 0.2 & 1.521 & [0.99,1] & 13.0 (414) & 37017 (3.6\%) & 403 & 386 & pulsar-like\\
	16.4 & $10^{-5}$ & 0.2 & 0.1484 & [0.99,1] & 13.0 (48) & 40280 (3.9\%) & 417 & 403 & pulsar-like\\
	16.4 & $10^{-6}$ & 0.2 & 1.484 &  [0.99,1] & 12.8 (109) & 39497 (3.9\%) & 416 & 403 & pulsar-like\\
\hline
	160 & $10^{-6}$  & 0.2 & 0.1521 & [0,1] & 4.46 (443) & 29864 (2.9\%) & 114 & 13 & hemisphere\\
	160 & $10^{-7}$ & 0.2 & 1.521 & [0,1] & 4.40 (427) & 29663 (2.9\%) & 114 & 11 & hemisphere\\
	16.4 & $10^{-5}$ & 0.2 & 0.1484 & [0,1] & 4.24 (341) & 29333 (2.9\%) & 110 & 8 & hemisphere\\
	16.4 & $10^{-6}$ & 0.2 & 1.484 & [0,1] & 4.05 (488) & 29456 (2.9\%) & 109 & 12 & hemisphere\\
\hline
	160 & $10^{-6}$  & 0.2 & 0.1521 & [0.1,0.8] & 3.22 (438) & 28965 (2.8\%) &17 & 0 & inter-cone\\
	160 & $10^{-7}$ & 0.2 & 1.521 & [0.1,0.8] & 3.79 (377) & 29030 (2.8\%) & 24 & 2 & inter-cone\\
	16.4 & $10^{-5}$ & 0.2 & 0.1484 & [0.1,0.8] & 2.90 (37) & 28630 (2.8\%) & 8 & 0 & inter-cone\\
	16.4 & $10^{-6}$ & 0.2 & 1.484 & [0.1,0.8] & 3.36 (168) & 28290 (2.8\%) & 13 & 0 & inter-cone\\
\hline
\hline
	160 & $10^{-6}$ & 0.44 & 0.1071 & [0.99,1] & 13.1 (352) & 16682 (1.6\%) & 181 & 164 & pulsar-like\\
	160 & $10^{-7}$ & 0.44 & 1.071 & [0.99,1] &  12.6 (32) & 16541 (1.6\%) & 182 & 169 & pulsar-like\\
	16.4 & $10^{-5}$ & 0.44 & 0.1045  & [0.99,1] & 12.9 (86) & 16747 (1.6\%) & 181 & 168 & pulsar-like\\
	16.4 & $10^{-6}$ & 0.44 & 1.045 &  [0.99,1] & 12.9 (431) & 16784 (1.6\%) &  179 & 166 & pulsar-like\\
\hline
	160 & $10^{-6}$ & 0.44 & 0.1071 & [0,1] & 5.07 (50) & 21334 (2.1\%)  & 326 & 157 & hemisphere\\
	160 & $10^{-7}$ & 0.44 & 1.071 & [0,1] & 4.85 (445) & 21158 (2.1\%) & 317 & 143 & hemisphere\\
	16.4 & $10^{-5}$ & 0.44 & 0.1045  & [0,1] & 4.80 (418) & 20322 (2.0\%) & 336 & 149 & hemisphere\\
	16.4 & $10^{-6}$ & 0.44 & 1.045 & [0,1] & 5.20 (187) & 20799 (2.0\%) & 343 & 177 & hemisphere\\
\hline
	160 & $10^{-6}$ & 0.44 & 0.1071 & [0.1,0.8] & 5.61 (390) & 22160 (2.2\%) & 271 & 179 & inter-cone\\
	160 & $10^{-7}$ & 0.44 & 1.071 & [0.1,0.8] & 5.65 (433) & 22335 (2.2\%) & 278 & 177 & inter-cone\\
	16.4 & $10^{-5}$ & 0.44 & 0.1045  & [0.1,0.8] & 5.46 (46)  & 21615 (2.1\%) & 299 & 185 & inter-cone\\
	16.4 & $10^{-6}$ & 0.44 & 1.045 & [0.1,0.8] & 5.37 (385) & 21268 (2.1\%) & 300 & 193 & inter-cone\\
\hline
\end{tabular}
 \label{tab:BM}
\end{table}


\subsection{Results}
\label{sec:results}

Given the large number of parameters, we have not done a systematic, complete exploration of the multidimensional phase space of models. However, we have 
explored numerous particular cases to look for trends related to the two phenomenological questions above. In doing this, we held one parameter not listed
in Table \ref{tab:BM} fixed for most runs: the initial value of the random number seed. Normally, this is Monte Carlo malpractice.  Two
models with identical initial random number seed and the same value of $n_{\rm bursts}$ and same ranges of $\costheta_b$ and $I_j(1)$ will start 
with identical sets of outbursts; that is
$\{\Phi_j,\bhat_j,I_j(1)\}$ will be the same. However, two models with the same $e^2$ but different values of $\Lambda$ will have different $\{\nhat_j\}$
hence different $\{\bhat_j\dotprod\nhat_j\}$  and different intensities $\{I_j\}$ so their subsets of detectable outbursts will be different. 
Models with different ranges of $\costheta_b$ and $I_j(1)$ start with identical sets of $\{\Phi_j\}$ which isolates the differences in properties
of detectable bursts associated with emitting geometry and precession. Here and there we
verified that the initial random number seed was not critical to qualitative features of the results.

For $n_{\rm bursts}$ in a total observing time $t_{\rm obs}=f_dn_{\rm p,cycle}P_p$, the burst rate is $n_{\rm bursts}/t_{\rm obs}$; for 
the simulations listed in Table \ref{tab:BM} 
$t_{\rm obs}=51.2$ d and the burst rate is $1024000/51.2{\,\rm d}=20,000\,{\rm d^{-1}}=0.231\,{\rm Hz}$. For our simulations, $\nustar\simeq 0.1$, $0.15$,
$1$ or $1.5$ Hz so the number of outbursts per spin period ranges from $\simeq 0.15$ to $\simeq 2.3$.
If all of these bursts were detectable, the spin frequency of the magnetar would be easy to find. In the simulation results, the burst detection
efficiency is {\sl at most} $\simeq 4\%$, which would correspond to {\sl at most} of order one burst per ten spin periods which, although not large, should still 
suffice to uncover the underlying spin period. The total number of outbursts was chosen so that the average number of {\sl detected} bursts 
per day would turn out to be $\sim 50$ in the simulations. This detection rate is comparable to the rate reported by \cite{2021arXiv210708205L} 
for about 50 days of
observations of FRB 121102. No convincing evidence for a spin frequency was found by \cite{2021arXiv210708205L}.

Table \ref{tab:BM} tabulates results for sixteen simulations. For all of these
\begin{enumerate}
\item $e^2=10$,
\item the distribution of intrinsic intensities ranges over a factor of 1000,
\item the beam width is $\theta_{\rm FWHM}=20^\circ$,
\item bursts are detectable over a range of observed intensities $I_{\rm max}/I_{\rm min}=1000$,
\item the distribution of beam directions is axisymmetric about
symmetry axis $\muhat=\ehat_3\cos\theta\mu+\sin\theta_\mu(\ehat_1\cos\varphi_\mu+\ehat_2\sin\varphi_\mu)$ with $(\theta_\mu,\varphi_\mu)=(30^\circ, 40^\circ)$,
\item there are daily observations lasting 0.1 d each over a total timespan of 512 d,
\item and the observer is at $i=52^\circ$ relative to the spin angular momentum of the star (Eq. (\ref{nhatdef})).
\end{enumerate}
The average number of bursts per day is $\sim 30-60$ for all of the tabulated models. Although the tabulated models are for $\epsd=0$ we do not regard this
as an essential limitation for two reasons. First, as ominscient simulators, we can nullify the effects of spindown entirely by adjusting the value of the
frequency tested from $\nustar$ to $\nu_{\star,{\rm d}}$; our code allows us to do this, and we have verified that $\Dhat_d(\nu_{\star, {\rm d}})$
with $\epsd\neq 0$ is virtually the same as $\Dhat(\nustar)$ with $\epsd=0$. Second, we do frequency searches on the most promising day with $\epsd\neq 0$
and detect frequency shifts for large enough spindown compared to our frequency resolution.

The table is divided into two categories, $\Lambda=0.2$ and $\Lambda=0.44$; more precisely $q^2=5/12$ for the upper half of the table and $q^2=12/5$ for
the lower half. Each half is subdivided into three parts that are distinguished by different ranges of beam directions:
\begin{enumerate}
\item ``pulsar-like'' models only allow beams in a very narrow cone of angles around $\muhat$, $0.99\leq\cos\theta_b\leq 1$;
\item  ``hemisphere'' models allow beams in any direction in the outward hemisphere relative to $\muhat$, $0\leq\cos\theta_b\leq 1$;
\item ``inter-cone'' models allow beams between the cones defined by $\cos\theta_{\rm min}=0.1$ and $\cos\theta_{\rm max}=0.8$ around $\muhat$.
\end{enumerate}
Inter-cone models exclude beams in a fairly narrow cone near the symmetry axis as well as beams orthogonal to it. A number of trends are apparent in
Table \ref{tab:BM}:
\begin{enumerate}
\item more bursts are detectable for $\Lambda=0.2$ than for $\Lambda=0.44$ in all cases;
\item for either value of $\Lambda$ all subcategories -- pulsar-like, hemisphere, intercone -- give similar results irrespective of the value of 
$\epsmag$;
\item $\nustar$ is readily detectable on $\sim 30\%$ of days for $\Lambda=0.44$ for pulsar-like, hemisphere and intercone geometries;
\item $\nustar$ is detectable on $\lesssim 3\%$ of all days for hemisphere models with $\Lambda=0.2$, but reproducible results for $\nustar$ (modulo spindown)
ought to emerge in a dedicated program of nearly daily observations that lasts long enough;
\item $\nustar$ is largely undetectable for inter-cone models with $\Lambda=0.2$.
\end{enumerate}
The uniformity of results within the various subcategories is not a complete surprise since $160/16.4$ is near ten so models with $P_p=160$ d and a given
value of $\epsmag$ are ought to resemble models with $P_p=16.4$ d and $10\epsmag$ closely. The dependence on $\epsmag$ for a given value of $P_p$ is weak.
The detection criteria in our simulations only depend on spin frequency implicitly via $\bhat\dotprod\nhat(\Phi)$, but this dependence is weak because
$\bhat$ varies widely and stochastically (except in the pulsar-like models). We expect that as long as the time between bursts is large compared with $1/\nustar$
final results should be insensitive to $\epsmag$.

Although we have not tabulated results for models in which beams can point in any direction (i.e. $-1\leq\cos\theta_b\leq 1$) we have simulated such
models; in general neither $\nustar$ nor $P_p$ is apparent in the results.

\begin{figure}[h]
\centering
\epsscale{0.9}
\plotone{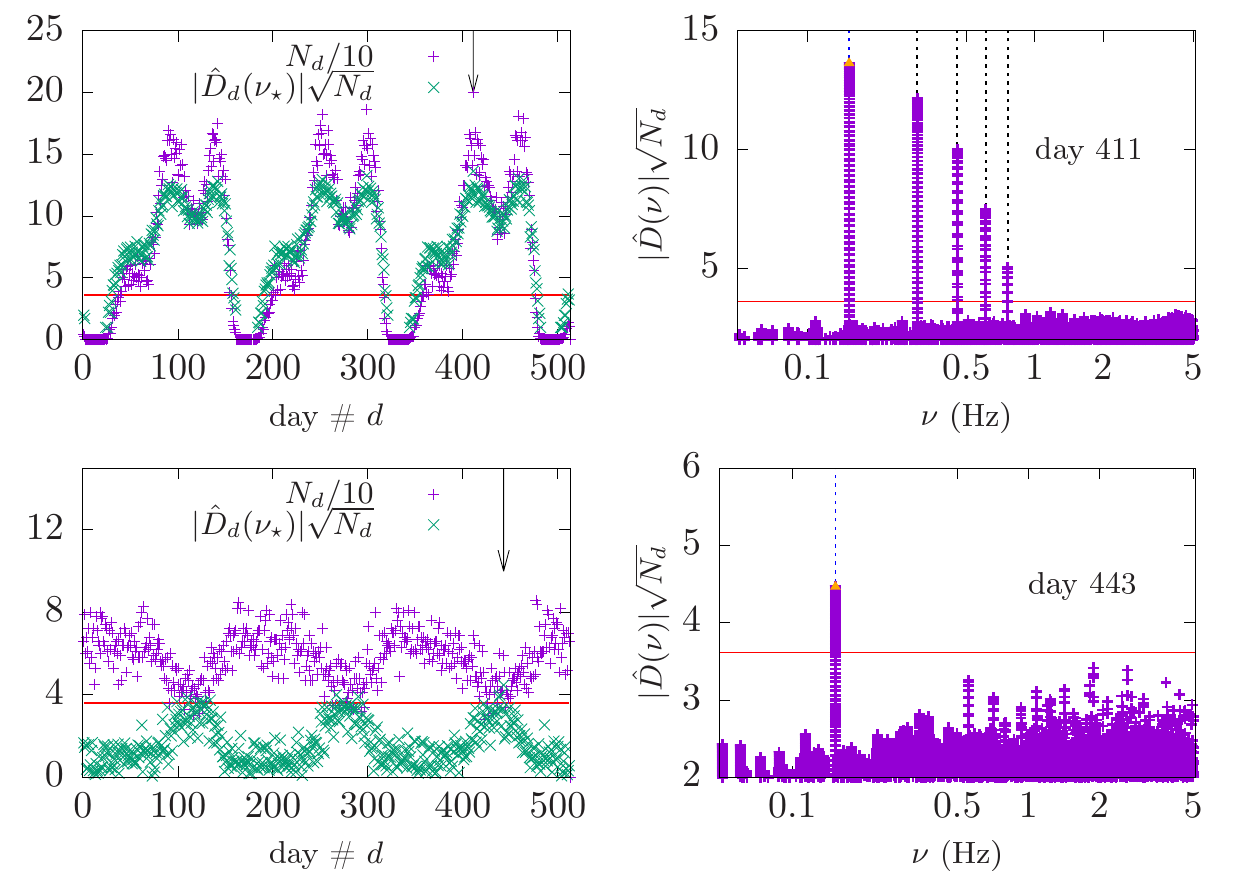}
\caption{Results for the most promising days for spin frequency detection for two models with $P_p=160$ d, $\Lambda=0.2$  and $\epsmag=10^{-6}$. A pulsar-like
model is shown in the top row and a hemisphere model in the bottom row. See Table \ref{tab:BM} for details.
Left panels show the number of bursts per day, $N_d$, divided by 10 (purple crosses)
and the value of $\vert{\hat D}_d(\nustar)\vert\sqrt{N_d}$ on each day (green x's). Downward pointing arrows indicate the day with largest value of 
$\vert{\hat D}_d(\nustar)\vert\sqrt{N_d}$, days 411 (top/pulsar-like) and 443 (bottom/hemisphere) respectively. Right panels show results of computing
$\vert{\hat D}_d(\nu)\vert\sqrt{N_d}$ on these days for frequencies $0.05\,{\rm Hz}\leq\nu\leq 5\,{\rm Hz}$ with equal logarithmic spacing $\Delta\nu/\nu
\simeq 10^{-5}$. Horizontal red lines are at $\vert{\hat D}_d(\nu)\vert\sqrt{N_d}=\sqrt{\ln N_{\rm freq}}\simeq 3.611$, the value above which about one
point ought to appear at random according to the Rayleigh distribution. For the models shown, detecting the spin frequency ought to be relatively easy:
the fundamental and four harmonics show up signficantly for pulsar like model but just the fundamental for hemisphere model.
}
\label{fig:variouspulsed}
\end{figure}

Fig. \ref{fig:variouspulsed} shows numerical results for two models where $\nustar$ ought to be detected. All of these results are for $P_p=160$ d,
$\epsmag=10^{-6}$ and $\Lambda=0.2$, so that $\nustar=0.1521$ Hz. The top panels are for the pulsar-like case and the bottom for the hemisphere case.
The left panels in both rows show $N_d/10$ for
each day (purple crosses); the precession cycle is evident in both panels. These panels also show $\Dhat_d(\nustar)\sqrt{N_d}$ for each day (green x's),
with the day on which $\Dhat_d(\nustar)\sqrt{N_d}$ is largest indicated by a downward arrow. The horizontal red lines in each figure are at 
$r_1(N_{\rm freq})=\sqrt{\ln N_{\rm freq}}\simeq 3.611$, which we take to be the threshold for detection of $\nustar$.
The right panels show the results of a period search on the most favorable day for detecting $\nustar$
using $N_{\rm freq}=460518$ test frequencies spaced at equal logarithmic intervals $\Delta\nu/\nu\simeq 10^{-5}$ between 0.05 Hz and 5 Hz. The value of
$\Dhat_d(\nustar)\sqrt{N_d}$ exactly at $\nustar$ is also shown as an orange triangle.  
The leftmost vertical dashed lines are at the spin frequencies for these two models;
for the pulsar-like model vertical dashed lines at four harmonics of $\nustar$ are also shown. The spin frequency and four harmonics are found easily
for the pulsar-like model; the spin frequency is also found for the hemisphere model.

\begin{figure}[ht]
\centering
\epsscale{0.9}
\plotone{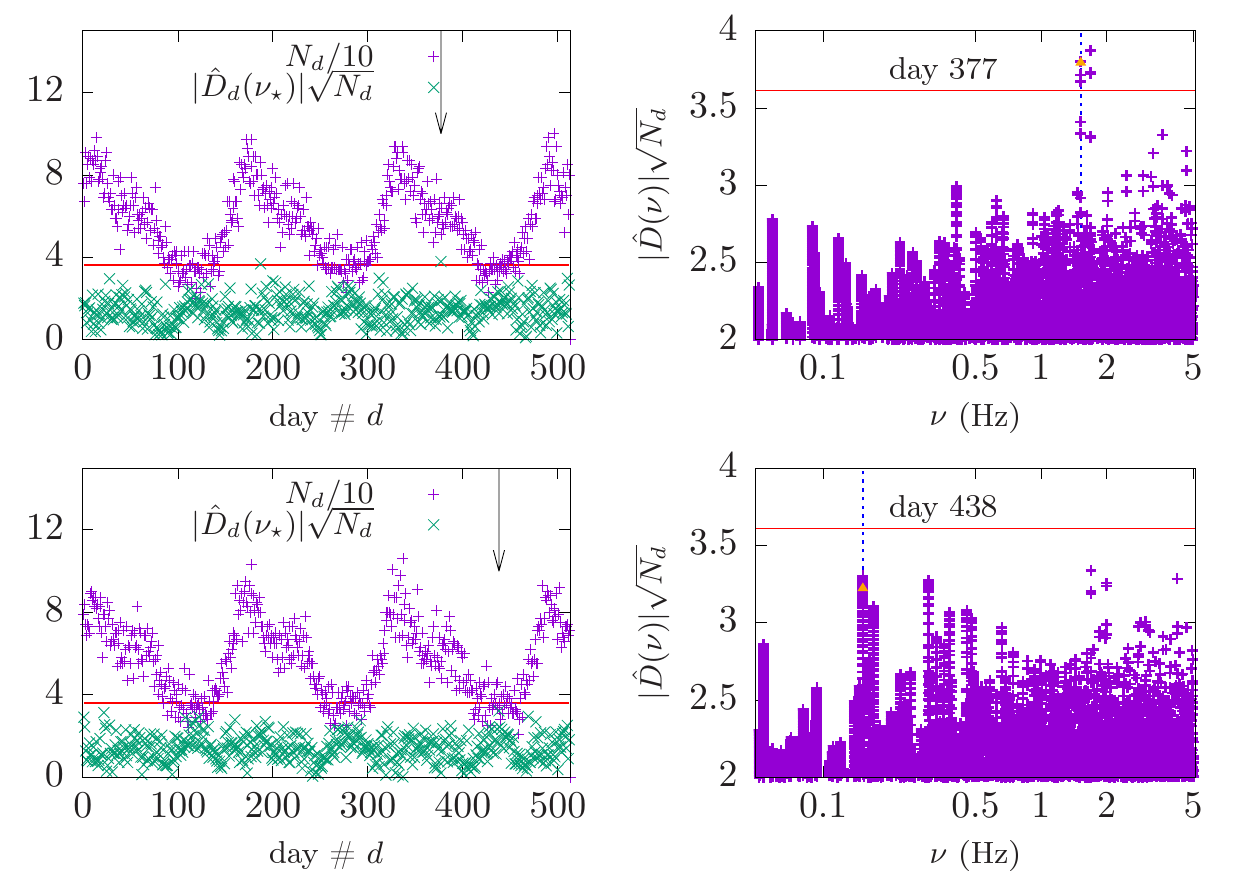}
\caption{Results for the most promising days for spin frequency detection for two inter-cone models with $P_p=160$ d, $\Lambda=0.2$ and $\epsmag=10^{-7}$ (top) and
$\epsmag=10^{-6}$ (bottom). The format is the same as in Fig. \ref{fig:variouspulsed}. For these models the spin frequency is not detectable unambiguously on
their respective most promising days.
}
\label{fig:various}
\end{figure}

Fig. \ref{fig:various} shows numerical results for two inter-cone models with $P_p=160$ d and $\epsmag=10^{-7}$ (top) and $\epsmag=10^{-6}$ (bottom).
For $\epsmag=10^{-7}$ the spin frequency should be detectable, whereas for $\epsmag=10^{-6}$ it is not. 
The spin frequency would be found significantly for $\epsmag=10^{-7}$, but a slighlty larger value of $\vert\Dhat(\nu)\vert\sqrt{N_d}$ is found around 1.7 Hz;.
This is not particularly troubling since both frequencies have values of $\vert\Dhat(\nu)\vert\sqrt{N_d}$ close to $r_1(N_{\rm freq})$, but it suggests that
$\nustar$ would not be detected on this particular day. (We reran this case with a different random number seed and found that $\vert\Dhat(\nu)\vert\sqrt{N_d}
<r_1(N_{\rm freq}$ on all days.) The spin frequency would not be found for $\epsmag=10^{-6}$. 

\begin{figure}[ht]
\centering
\epsscale{0.9}
\plotone{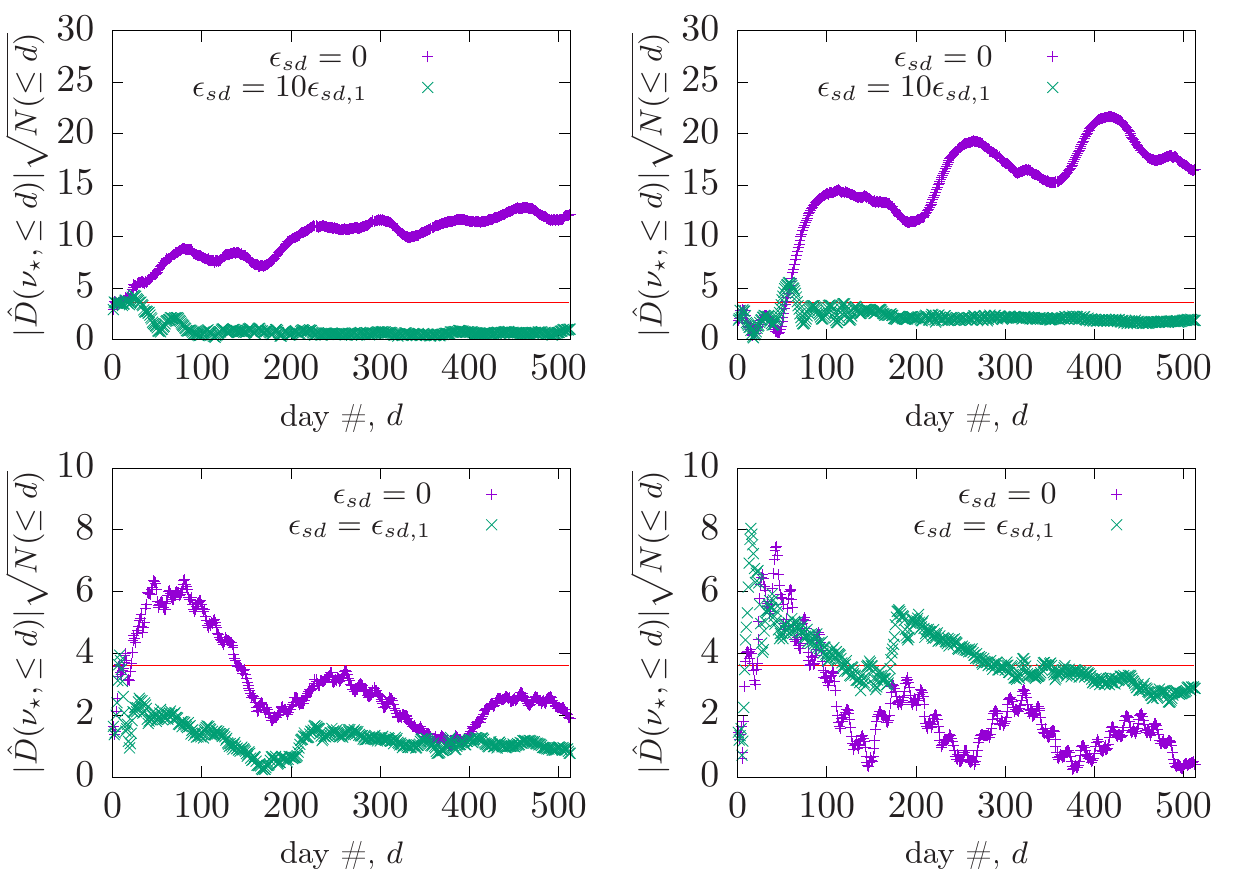}
\caption{Cumulative $\vert\hat D(\nustar,\leq d)\vert\sqrt{N(\leq d)}$ as a function of day number $d$. Top: $P_p=160$ d, 
$\epsmag=10^{-6}$ nested-cone models with $\Lambda=0.2$
(left) and $\Lambda=0.44$ (right), and $\epsd=0$ (purple crosses) and $\epsd=10\epsilon_{sd,1}$ (green x's). Bottom: $P_p=16.4$ d, $\epsmag=10^{-5}$ with
$\Lambda=0.2$ (left) and $\Lambda=0.44$ (right), and $\epsd=0$ (purple crosses) and $\epsd=\epsilon_{sd,1}$ (green x's).
}
\label{fig:Dhacc}
\end{figure}

So far, we have concentrated on searching for $\nustar$ on individual days, presenting frequency spectra only on the most promising days for detection. 
Alternatively, the frequency search can be performed on the entire data set. Fig \ref{fig:Dhacc} shows results for the cumulative value 
$\vert\Dhat(\nustar,\leq d)\vert$
computed by performing the sum Eq. (\ref{Ddef}) over the $N(\leq d)$ bursts detected up to the end of day $d$ and multiplying by $\sqrt{N(\leq d)}$. All four
panels in Fig. \ref{fig:Dhacc} are computed for nested-cone geometry with $0.1\leq\cos\theta_b\leq 0.8$. The top panels are for $P_p=160$ d and $\epsmag=10^{-6}$
and the bottom for $P_p=16.4$ d and $\epsmag=10^{-5}$ so spin frequencies are comparable in all cases. The left panels are for $\Lambda=0.2$, where Table \ref{tab:BM}
indicates no promising days for burst detections, and the right panels are for $\Lambda=0.44$, for which we expect $\approx 200$ promising days. The top left
panel shows that the value of $\vert\Dhat(\nustar,\leq d)\vert\sqrt{N(\leq d)}$ 
generally increases with $d$ for $\epsd=0$, apart from fluctuations, suggesting that detection
may be possible in a cumulative analysis. However, Eq. (\ref{epssdoverepsdone}) implies that $\epsd/\epsilon_{{\rm sd},1}\approx 17B_{D,14}^2$
for the top left panel in Fig. \ref{fig:Dhacc} and $\approx 6B_{D,14}^2$ for the right panel. At these levels, the accumulated spindown over many days of observation
becomes important, and discovering $\nustar$ from a cumulative analysis that neglects spindown is problematic. The situation for $P_p=16.4$ d appears to be 
more complicated. As the left panel shows, although $\vert\Dhat(\nustar,\leq d)\vert\sqrt{N(\leq d)}$ increases at first for $\epsd=0$, ultimately it decreases while 
fluctuating considerably; the same sort of behavior is evident in the right panel. For these cases, $\epsd\approx 0.17B_{D,14}^2\epsilon_{{\rm sd},1}$ 
and $0.06B_{D,14}^2\epsilon_{{\rm sd},1}$, respectively, so spindown is less important for $B_{D,14}=1$. We show what happens for $\epsd=\epsilon_{{\rm sd},1}$ 
in both panels: spindown this fast further suppresses accumulation of evidence for $\nustar$ in the left panel, but actually can enhance it, at least for awhile, 
in the case depicted in the right panel.

\newpage
\section{Conclusions}
\label{sec:conc}

The first part of this paper has been devoted to the theory of precession of neutrons stars whose distortions are due to strong internal magnetic fields.
We have stressed that unless internal field strengths are very large precession ought to be frustrated by pinning of neutron superfluid vortices, to nuclei in the
crust \citep{1977ApJ...214..251S} and to flux tubes associated with Type II proton superconductivity in the core \citep{2003PhRvL..91j1101L}. We argue that 
internal magnetic fields with strength $\gtrsim 10^{16}\,\Gauss$ must pervade the star if precession is to be possible. As we have discussed, fields this strong
can prevent proton superconductivity in the core (see Eq. (\ref{Hctwo}) and associated discussion), are strong enough to shatter the crust and may even prevent 
or at least substantially alter neutron superfluidity in both core and crust (see Eq. (\ref{BCC}) and associated discussion). Avoiding vortex line pinning is
a very stringent requirement:
even very localized regions where neutron vortices pin can prevent slow precession if they comprise just a modest fraction -- say 0.1\% -- of the moment of
inertia of the star \citep{1977ApJ...214..251S}. 

These considerations led us to examine implications of a three component model of the magnetic field in magnetars: (1) a dipole component with characteristic strength
$\sim 10^{14}\,\Gauss$; (2) a toroidal component with characteristic strength $\sim 10^{15}-10^{16}\,\Gauss$; and (3) a disordered field with 
characteristic strength $\sim 10^{16}\,\Gauss$. Because the volume occupied by the toroidal field may be limited (as in the models of \cite{2013MNRAS.433.2445A})
the disordered component may be critical for suppressing superfluidity and superconductivity, which is a necessary condition for slow precession. Since
small-scale, disordered fields may decay in a timescale $\sim 1000$ years (see Eq. (\ref{tambi}) and \cite{1992ApJ...395..240R},
\cite{2011MNRAS.413.2021G}, \cite{2017MNRAS.465.3416P} and \cite{2017PhRvD..96j3012G}) a magnetar may only precess for a relatively short time, until the
disordered field that enables slow precession decays significantly. The decay of the disordered field may also end the bursting phase in
the life of a magnetar, for this component contains most of the stellar magnetic energy and
is capable of stimulating internal fluctuations that propagate into the magnetosphere, where they might trigger burst activity.

Illustrative but schematic models for magnetic distortion in \S\ref{sec:magprec} imply that the resulting quadurpolar deformation is almost certainly triaxial and 
probably prolate.  We developed the theory of triaxial precession in \S\ref{sec:triaxprec}, noting in particular that large amplitude precession can be excited
as a result of small shearing motions involving only tiny fractions of the magnetic energy of the star. We included spindown in the theory developed in
this section (using the spindown formula found by \cite{2012ApJ...746...60L}): \S\ref{sec:timingmodel} develops a timing model $t(\Phi)$ that relates 
clock time to precession phase when spindown is included, and \S\ref{sec:long_term_evolution} develops the secular effects of spindown on precession dynamics
for the triaxial case, a generalization of \cite{1970ApJ...160L..11G} which dealt with axisymmetric, oblate precession. We have yet to explore possible phenomenological
implications of the secular evolution. The timing model exhibits the expected systematic spindown, as altered by precession, but
also includes important cyclical terms that vary periodically with precession (discussed less generally by \cite{1993ASPC...36...43C}). 

In the introduction, we asked whether the spin frequency ought to be detected for a magnetar precessing with a known precesion period. So far, no spin frequency
is apparent in either FRB 121102 or FRB 180916.J0158+65. This may be simply because we have not detected enough bursts from these FRBs to find evidence for
their spin periods, or it may be that doing so is virtually impossible because of physical properties of these objects and the FRB mechanism. If the reason we
have yet to detect spin frequencies is that we need more burst detections, how sensitive and systematic must an observing program be to find the spin convincingly ?

In order to address this issue we constructed a specific stochastic model for FRBs in \S\ref{apptofrbs}. In this model, FRBs are associated with outbursts that
occur randomly in time with energy output that is beamed into a range of directions that we select. Generally, we confine the beam directions to be outward
relative to a reference (magnetic) axis; for beam directions that are totally random neither the spin frequency nor the precession period is discernible.
Thus, the existence of repeated precession cycles for FRB 121102 and FRB 180916.J0158+65 already shows that they are not caused by beamed emission directed 
entirely at random.

Our model offers some hope for detecting spin frequencies, 
as precession 
implies that there is a bias that favors detection of optimally directed beams. The analytic model in Appendix \ref{app:analytic} demonstrates that the
dependence on both the spin and precession frequencies arise from the motion of the unit vector to the observer in the rotating frame of reference, but the
dependences may be very weak.
In order to assess whether or not the spin frequency can be detected, we computed
$r_d(\nustar)=\vert\Dhat_d(\nustar)\vert\sqrt{N_d}$ for each day $d$ in our hypothetical observing program; $N_d$ is the number of bursts detected on day $d$ 
and $\Dhat(\nu)$ is  defined in Eq. (\ref{Ddef}). For a frequency search with $N_{\rm freq}=460518$ frequencies spanning the range 
$0.05-5$ Hz with equal logarithmic spacing, $\Delta\nu/\nu=10^{-5}$, on any given day, the largest value that should arise at random is approximately 
$r_1(N_{\rm freq})=\sqrt{\ln N_{\rm freq})}\approx 3.611$. 
The spin frequency ought to be detectable on days when $r_d(\nustar)>r_1(N_{\rm freq})$. Finding $\nustar$ is likelier for cases where the number of days
with $r_d(\nustar)>r_1(N_{\rm freq})$ is a substantial fraction of the total number of days on which observations are done.

Although we have only computed a modest number of models, the results reported in Table \ref{tab:BM} divide qualitatively into two classes depending on the 
value of $q^2=e^2\Lambda^2/(1-\Lambda^2)$. Based on the criterion described above, we believe that detecting the burst frequency is likely when $q^2>1$ irrespective
of the value of $\epsmag$ or the distribution of beam directions. However, the situation for $q^2<1$ is more complicated. Although detecting $\nustar$ ought to 
be easy for pulsar-like models, where the range of beam directions relative to the reference axis is small, widening this range diminishes the odds
of detection considerably. Allowing beam directions anywhere in the outward hemisphere relative to the axis would lead to detections on $\lesssim 3\%$ of the
days during which observations are done. But, restricting beams to avoid directions moderately close to the axis and perpendicular to it makes detecting
the spin frequency is nearly impossible in our models.

Finally, the simulations all indicate that the fraction of outbursts that are ultimately detectable is small: the largest fraction of all outbursts that were
detectable in our models was 3.9\%, for pulsar-like models with $q^2<1$,  and is at most $2.9\%$ for all other models we have simulated. 
That means that the model simulated here is not very energy-efficient, in that
at least $\sim 25-50$ times as much energy is being emitted in FRBs than we would deduce from observations. One might expect that beaming mitigates the energetic
requirements, and of course for a given peak intensity the total emitted intensity is $\propto 1/\kappa\sim\theta_{\rm FWHM}^2$. 
Our calculations only cover a single Gaussian beam width, $\theta_{\rm FWHM}=20^\circ$. Lowering $\theta_{\rm FWHM}$ reduces the total number of detections
at fixed outburst rate, which we have found to be roughly $\propto\theta_{\rm FWHM}^2$ via sporadic exploration of the phase space. Assuming this to be 
true, the overall amount of energy required in the stochastic model would be roughly independent of $\theta_{\rm FWHM}$ for small values of the beam width: 
the total emission per beam is $\propto\theta_{\rm FWHM}^2$ but the number of undetected beams per detected beam is $\propto 1/\theta_{\rm FWHM}^2$.
Moreover, with fewer burst detections per day uncovering the magnetar spin frequency becomes harder.

\begin{appendix}

\section{Useful Integrals and Details of Computing the Timing Model}
\label{app:funcs}

\begin{table}[h]
\caption{Useful Integrals and Averages}
\centering
\begin{tabular}{|c|c|}
\hline
	Integral\footnote{$Q=q$ for $q<1$ and $Q=1/q$ for $q>1$.}        &  Result \\
\hline
\hline
	$\int_0^{\Phi}d\Phi'\dn(\Phi')$ & $\varphi(\Phi)$\\
	$\int_0^\Phi d\Phi'\cn\Phi'\dn\Phi'$ & $\sn\Phi$\\
	$\int_0^\Phi d\Phi'\sn\Phi'\dn\Phi'$ & $-\cn\Phi$\\
	$\int_0^\Phi d\Phi'\cn\Phi'=\int_0^{\varphi(\Phi)}\frac{d\varphi\cos\varphi}{\sqrt{1-Q^2\sin^2\varphi}}$ & $\frac{\asin[Q\sin\varphi(\Phi)]}{Q}$
	\footnote{$-\pi/2\leq\asin z\leq \pi/2$.}\\
	$\int_0^\Phi d\Phi'\sn\Phi'=\int_0^{\varphi(\Phi)}\frac{d\varphi\sin\varphi}{\sqrt{1-Q^2+Q^2\cos^2\varphi}}$ & $\frac{1}{Q}\ln\left[\frac{1+Q}{Q\cn\Phi
	+\sqrt{Q^2\cn^2\Phi+1-Q^2}}\right]=\frac{1}{Q}\ln\left[\frac{\dn(\Phi)-Q\cn(\Phi)}{1-Q}\right]$\\
	$\int_0^\Phi d\Phi'\sn^2\Phi'$ & $\frac{1}{Q^2}\left[F(\varphi(\Phi)|Q)-E(\varphi(\Phi)|Q)\right]$\\
	& $E(\varphi|Q)=\int_0^\varphi d\varphi'\sqrt{1-Q^2\sin^2\varphi'}$\\
	$\int_0^\Phi d\Phi'\sn\Phi'\cn\Phi'$ & $-\frac{\dn\Phi}{Q^2}$\\
	$\langle\sn^2\Phi\rangle$ & $\frac{1}{Q^2}\left[1-\frac{E(\pi/2|Q)}{F(\pi/2|Q)}\right]$\\
	$\langle\dn\Phi\rangle$ & $\frac{\pi/2}{F(\pi/2|Q)}$\\
	$\langle\frac{1}{\dn\Phi}\rangle$ & $\frac{\pi/2}{F(\pi/2|Q)\sqrt{1-Q^2}}$\\
	$\left\langle\frac{\sn^2\Phi}{\dn\Phi}\right\rangle$ & $\frac{\pi/2}{Q^2F(\pi/2|Q)}\left(\frac{1}{\sqrt{1-Q^2}}-1\right)$\\
	$\left\langle\frac{\cn^2\Phi}{\dn\Phi}\right\rangle$ & $\frac{\pi/2}{Q^2F(\pi/2|Q)}\left(1-\sqrt{1-Q^2}\right)$\\
\hline
\end{tabular}
\label{tab:usefulintegrals}
\end{table}
Using Table \ref{tab:usefulintegrals} we find
\ba
& &\int_0^\Phi d\Phi'\int_0^{\Phi'}d\Phi''\cn\Phi''\sn\Phi''=\frac{1}{Q^2}\int_0^\Phi d\Phi'(1-\dn\Phi')=\frac{\Phi-\varphi(\Phi)}{Q^2}
=\frac{\Phi(1-\langle\dn\phi\rangle)}{Q^2}+\frac{\Phi\langle\dn\Phi\rangle-\varphi(\Phi)}{Q^2}
\nonumber\\& &~~~~~~~~~~~~~~~~~
\equiv \frac{\Phi(1-\langle\dn\phi\rangle)}{Q^2}+C_2(\Phi|Q)
\nonumber\\
& &\int_0^\Phi d\Phi'\int_0^{\Phi'}d\Phi''\cn\Phi''\dn\Phi''=\int_0^\Phi d\Phi'\sn\Phi'=\frac{1}{Q}\ln\left[\frac{\dn(\Phi)-Q\cn(\Phi)}{1-Q}\right]
\equiv C_3(\Phi|Q)
\nonumber\\
& &\int_0^\Phi d\Phi'\int_0^{\Phi'}d\Phi'\sn\Phi''\dn\Phi''=\int_0^\Phi d\Phi'(1-\cn\Phi')=\Phi-\frac{\asin[Q\sin\varphi(\Phi)]}{Q}
\equiv\Phi-C_4(\Phi|Q)
\ea
where we rewrote the first integral to isolate the secular term from the strictly periodic one. The third integral also has a secular term. These
terms are $\propto\Phi\propto\phi$ and, in effect, renormalize the initial spin period.

Finally, we consider terms $\propto\sn^2\Phi$; we clearly need to remove $\snsqav$, which produces a term in $t(\phi)$ that is $\propto\phi^2$.
We assume that what remains is periodic, so we focus just on the period starting at $\Phi=0$. We then find
\ba
& &C_1(\Phi|Q)\equiv\int_0^\Phi d\Phi'\int_0^{\Phi'}d\Phi''\left(\sn^2\Phi''-\snsqav\right)=\int_0^\Phi\frac{d\Phi'[F(\varphi(\Phi')|Q)-E(\varphi(\Phi')|Q)]}{Q^2}
-\frac{\snsqav\Phi^2}{2}
\nonumber\\
& &=\frac{\Phi^2}{2}\left(\frac{1}{Q^2}-\snsqav\right)-\frac{1}{q^2}\int_0^\Phi d\phi'E(\varphi(\Phi')=\frac{\Phi^2E(\pi/2|Q)}{2Q^2F(\pi/2|Q)}
-\frac{1}{Q^2}\int_0^\Phi d\Phi'E(\varphi(\Phi')|Q)~.
\ea
Available routines for evaluating complete elliptic functions return values for $\varphi(\Phi)\leq\pi/2$ \citep{2002nrca.book.....P} which covers
all of the values of these functions; for $\pi/2<\varphi\leq\pi$ we substitute
\be
E(\varphi|Q)=E(\pi|Q)-E(\pi-\varphi|Q)=2E(\pi/2|Q)-E(\pi-\varphi|Q)
\ee
to get
\ba
& &\int_0^\Phi d\Phi'E(\Phi'|Q)=\int_0^{\pi/2}\frac{d\varphi E(\varphi|Q)}{\sqrt{1-q^2\sin^2\varphi}}+2E(\pi/2|Q)(\Phi-F(\pi/2|Q))
-\int_{\pi/2}^{\varphi(\Phi)}\frac{d\varphi'E(\pi-\varphi'|Q)}{\sqrt{1-Q^2\sin^2\varphi'}}
\nonumber\\& &
=\int_0^{\pi/2}\frac{d\varphi E(\varphi|Q)}{\sqrt{1-q^2\sin^2\varphi}}+2E(\pi/2|Q)(\Phi-F(\pi/2|Q))
-\int_{\pi-\varphi(\Phi)}^{\pi/2}\frac{d\psi E(\psi|Q)}{\sqrt{1-Q^2\sin^2\psi}}
\nonumber\\& &
=2E(\pi/2|Q)(\Phi-F(\pi/2|Q))+\int_0^{\pi-\varphi(\Phi)}\frac{d\psi E(\psi|Q)}{\sqrt{1-Q^2\sin^2\psi}}~.
\ea
$C_1(\Phi|Q)=0$ at $\Phi=0$ and $\Phi=2F(\pi/2|Q)$, and has its peak value at $\Phi=F(\pi/2|Q)$. 

\begin{figure}[ht]
\centering
\epsscale{0.8}
\plotone{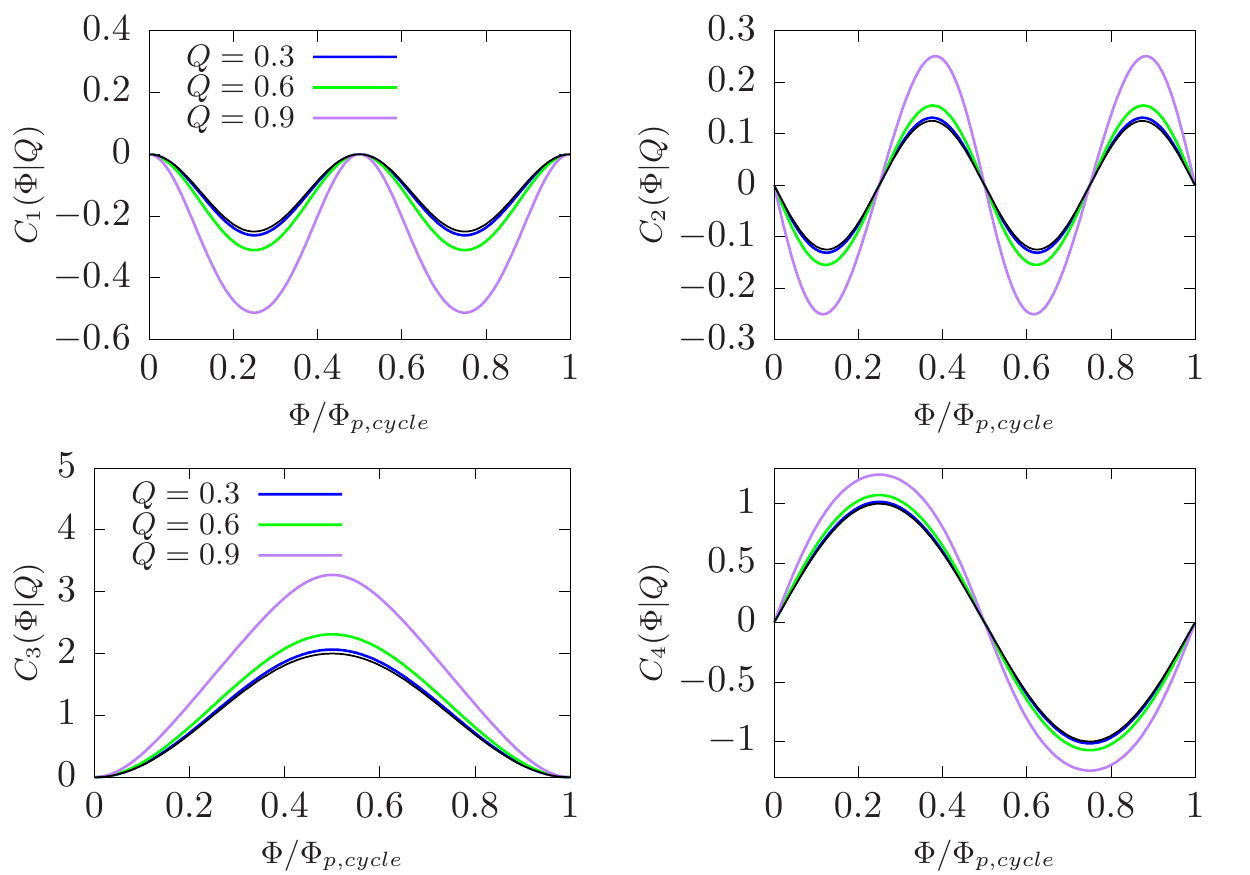}
	\caption{$C_i(\Phi|Q)$ versus $\Phi$ for one precession cycle and $Q=0$, $0.3$, $0.6$ and $0.9$.
}
\label{fig:precessandtimeplot}
\end{figure}

Fig. \ref{fig:precessandtimeplot} shows results for $C_i(\Phi|Q)$ for one precession cycle for $Q=0.3$, $0.6$ and $0.9$, and also for $Q=0$ (thin black line), 
for which the limiting forms are 
\be
C_1(\Phi)=\frac{\cos 2\Phi -1}{8}
~~C_2(\Phi)=-\frac{\sin 2\Phi}{8}
~~
C_3(\Phi)=1-\cos\Phi
~~
C_4(\Phi)=\sin\Phi~.
\label{Cizeroq}
\ee
The functions $C_1(\Phi|q)$ and $C_2(\Phi|q)$ have periods equal to half of the precession period. The functions $C_3(\Phi|q)$ and $C_4(\Phi|q)$ have
periods equal to a full precession cycle.
Notice that although there is no secular variation of $C_3(\Phi|q)$ this function has a nonzero mean over its full cycle of variation, which would
manifest itself as a offset in $t(\phi)$.

\section{Analytic Theory of Detection Probability}
\label{app:analytic}

\setcounter{equation}{0}

Eq. (\ref{Ijmodel}) relates the intrinsic intensity $I$ and the peak intensity $I(1)$; in our model a burst is detectable if $I>I_{\rm min}$. The probability
that a burst with peak intensity $I(1)$ is detectable at precession phase $\Phi$ is
$$
P_{\rm detect}(I(1),\Phi)=\int d^2\bhat\,n(\bhat)\,\Theta(I-I_{\rm min})
=\int d^2\bhat\,n(\bhat)\,\Theta\left(I(1)\exp[\kappa(\bhat\dotprod\nhat(\Phi)-1)]-I_{\rm min}\right)~,
$$
where $n(\bhat)$ is the distribution of beam directions (normalized to one) and $\Theta(\cdots)$ is the Heaviside function;
use
\ba
& &\Theta\left(I(1)\exp[\kappa(\bhat\dotprod\nhat(\Phi)-1)]-I_{\rm min}\right)=\Theta((I(1)/I_{\rm min})\exp[\kappa(\bhat\dotprod\nhat(\Phi)-1)]-1)
\nonumber\\& &~~~~~~~~~~
=\Theta(\ln(I(1)/I_{\rm min})+\kappa[\bhat\dotprod\nhat(\Phi)-1])=\Theta(\kappa^{-1}\ln(I(1)/I_{\rm min})-1+\bhat\dotprod\nhat(\Phi))
\nonumber
\ea
to rewrite as
$$
P_{\rm detect}=\int d^2\bhat\,n(\bhat)\,\Theta(\kappa^{-1}\ln(I(1)/I_{\rm min})-1+\bhat\dotprod\nhat(\Phi))~.
$$
In a right handed coordinate system defined by unit vectors $\ehat_a,\ehat_b,\muhat$ let
$$
\nhat(\Phi)=\muhat\cos\tnh+\sin\tnh(\ehat_a\cos\phinh+\ehat_b\sin\phinh)~,
$$
so that
$$
\bhat\dotprod\nhat(\Phi)=\cos\theta_b\cos\tnh+\sin\theta_b\sin\tnh\cos(\varphi_b-\phinh)\equiv\cos\theta_b\cos\tnh+\sin\theta_b\sin\tnh\cos\tilde\varphi_b~;
$$
then the Heaviside function requires that
$$
1\geq\cos\theta_b\cos\tnh+\sin\theta_b\sin\tnh\cos\tilde\varphi_b>1-\frac{\ln(I(1)/I_{\rm min})}{\kappa}
~,
$$
and therefore $\bhat\dotprod\nhat(\Phi)$ must be close to one for very large values of $\kappa$. Thus if we define $\theta_b=\tnh+\delta_b$ where $\delta_b\ll 1$
and assume that $\tilde\varphi_b\ll 1$
$$
\cos\theta_b\cos\tnh+\sin\theta_b\sin\tnh\cos\tilde\varphi_b=\cos(\theta_b-\tnh)+\sin\theta_b\sin\tnh(\cos\tilde\varphi_b-1)
\approx 1-\onehalf\left(\delta_b^2+\sin^2\tnh\tilde\varphi^2_b\right)
$$
and the Heaviside function requires that
$$
1\geq 1-\onehalf\left(\delta_b^2+\sin^2\tnh\tilde\varphi^2_b\right)>1-\frac{\ln(I(1)/I_{\rm min})}{\kappa}~\Rightarrow~
0\leq\delta_b^2+\sin^2\tnh\tilde\varphi^2_b<\frac{2\ln(I(1)/I_{\rm min})}{\kappa}\equiv R^2~.
$$
We assume that $n(\bhat)=\ntil(\cos\theta_b)/2\pi$ is only a function of $\cos\theta_b$; with this assumption
$$
d^2\bhat n(\bhat)=\frac{d\varphi}{2\pi}\,d\cos\theta_b\,\ntil(\cos\theta_b)
\simeq\frac{d\tilde\varphi_b}{2\pi}\,d\delta_b\sin\tnh\,\ntil(\cos\tnh)
$$
With these substitutions we find
\ba
& &P_{\rm detect}(I(1),\Phi)\simeq\int_{-R}^{+R}d\delta_b\,\sin\tnh\,\ntil(\cos\tnh)
\int_{-\sqrt{R^2-\delta_b^2}/\sin\tnh}^{+\sqrt{R^2-\delta_b^2}/\sin\tnh}\frac{d\tilde\varphi_b}{2\pi}
\nonumber\\& &~~~~~~~
=\frac{2R^2\ntil(\cos\tnh)}{\pi}\int_0^1 dx\,\sqrt{1-x^2}
=\frac{\ntil(\cos\tnh)\ln(I(1)/I_{\rm min})}{\kappa}~.
\label{Pdetect}
\ea
\begin{enumerate}
\item $P_{\rm detect}$ increases with increasing $I(1)$ weakly, decreases $\propto 1/\kappa$ 
as $\kappa$ increases, and is only nonzero where $n(\cos\tnh)\neq 0$.
\item If $\ntil(\cos\theta_b)$ is uniform, as would be the case if beam directions are random and isotropic, then there is no $\Phi$ dependence
so there is no imprint of either the spin frequency or precession frequency on $P_{\rm detect}$. 
\item $\Phi$ dependence arises from nonumiformity in $\ntil(\cos\theta_b)$; in our models, nonuniformity is a consequence of boundaries in the distribution
of beam directions. 
\end{enumerate}

Consider the region near $\theta_{\rm max}$. If $\theta_{\rm max}-R<\tnh<\theta_{\rm max}$ and $\ntil(\cos\theta_b)=0$ at $\theta_b>\theta_{\rm max}$ then
\be
P_{\rm detect}(I(1),\Phi)\simeq\frac{\ntil R}{\pi}\int_{-R}^{\theta_{\rm max}-\tnh}d\delta_b\,\sqrt{1-\delta_b^2/R^2}
=\frac{\ntil R^2}{\pi}\left[\frac{\pi}{4}+\frac{1}{2}\left(\sin^{-1}\Delta+\Delta\sqrt{1-\Delta^2}\right)\right]
\label{Pdetectinside}
\ee
where $\Delta=(\theta_{\rm max}-\tnh)/R<1$ and $\ntil$ is the uniform value inside the region containing beams; 
if $\tnh>\theta_{\rm max}$ then $-R<\delta_b<\theta_{\rm max}-\tnh<0$ and 
\be
P_{\rm detect}(I(1),\Phi)=\frac{\ntil R}{\pi}\int_{-R}^{\theta_{\rm max}-\tnh}d\delta_b\sqrt{1-\delta_b^2/R^2}=\frac{\ntil R^2}{\pi}\left[\frac{\pi}{4}
-\frac{1}{2}\left(\sin^{-1}\vert\Delta\vert+\vert\Delta\vert\sqrt{1-\Delta^2}\right)\right]~.
\label{Pdetectoutside}
\ee
Eq. (\ref{Pdetectinside}) can be used for $\tnh$ inside or outside provided that we use $\sin\Delta=-\sin\vert\Delta\vert$ for $\Delta<0$. Near $\theta_{\rm min}$,
similar considerations imply that
$-1<\delta_b<\tnh-\theta_{\rm min}$; then we get the same results but with $\Delta=(\tnh-\theta_{\rm min})/R$. For $\theta_{\rm min}=0$, $0\leq\delta_b\leq\tnh$
($0\leq\Delta\leq 1$). Otherwise, $P_{\rm detect}=0$ for $\tnh\leq\theta_{\rm min}-R$ and for $\tnh>\theta_{\rm max}+R$.
These results depend on $\Phi$ via $\Delta$ and
imprint information about both the spin frequency and the precession cycle on our models. We also note that the range of values $\tnh$ is model dependent via
$\Lambda$, $e^2$, and the rate at which outbursts occur, which may depend on $\Phi$ (but doesn't in our models).
If $R(\Phi,I(1))dI(1)d\Phi$ is the rate of outbursts with intrinsic intensity in $I(1)+dI(1)$ then
\be
dr_{\rm detect}(\Phi)=d\Phi\int dI(1)R(I(1),\Phi)P_{\rm detect}(I(1),\Phi)
\ee
is the rate of detection of bursts in $d\Phi$.

These results can be generalized to beams distributed about multiple axes by replacing
$$
n(\bhat)\to\sum_jp_jn_j(\bhat)
$$
where $p_j$ is the probability that a beam comes from the distribution around the axis $\muhat_j$ and $n_j(\bhat)$ is normalized to one.
The average beam direction is then the sum of $p_j\langle\bhat_j\rangle$.

\end{appendix}

\begin{acknowledgments}
SC and JMC acknowledge support from the National Science Foundation (NSF) under AAG award 1815242 and are members of the NANOGrav Physics Frontiers Center supported by  NSF award  1430284. IW thanks Jeevak Parpia, Armen Sedrakian and Peter Rau for helpful comments.
\end{acknowledgments}

\bibliographystyle{hapj}
\bibliography{precessionsection}

\begin{thebibliography}{78}
\expandafter\ifx\csname natexlab\endcsname\relax\def\natexlab#1{#1}\fi

\bibitem[{{Abramowitz} \& {Stegun}(1972)}]{1972hmfw.book.....A}
{Abramowitz}, M., \& {Stegun}, I.~A. 1972, {Handbook of Mathematical Functions}

\bibitem[{{Akg{\"u}n} {et~al.}(2006){Akg{\"u}n}, {Link}, \&
  {Wasserman}}]{2006MNRAS.365..653A}
{Akg{\"u}n}, T., {Link}, B., \& {Wasserman}, I. 2006, \mnras, 365, 653,
  astro-ph/0506606

\bibitem[{{Akg{\"u}n} {et~al.}(2013){Akg{\"u}n}, {Reisenegger}, {Mastrano}, \&
  {Marchant}}]{2013MNRAS.433.2445A}
{Akg{\"u}n}, T., {Reisenegger}, A., {Mastrano}, A., \& {Marchant}, P. 2013,
  \mnras, 433, 2445, 1302.0273

\bibitem[{{Akg{\"u}n} \& {Wasserman}(2008)}]{2008MNRAS.383.1551A}
{Akg{\"u}n}, T., \& {Wasserman}, I. 2008, \mnras, 383, 1551, 0705.2195

\bibitem[{{Alpar} {et~al.}(1984){Alpar}, {Pines}, {Anderson}, \&
  {Shaham}}]{1984ApJ...276..325A}
{Alpar}, M.~A., {Pines}, D., {Anderson}, P.~W., \& {Shaham}, J. 1984, \apj,
  276, 325

\bibitem[{{Anderson} {et~al.}(1982){Anderson}, {Alpar}, {Pines}, \&
  {Shaham}}]{1982PMagA..45..227A}
{Anderson}, P.~W., {Alpar}, M.~A., {Pines}, D., \& {Shaham}, J. 1982,
  Philosophical Magazine, Part A, 45, 227

\bibitem[{{Anderson} \& {Itoh}(1975)}]{1975Natur.256...25A}
{Anderson}, P.~W., \& {Itoh}, N. 1975, Nature, 256, 25

\bibitem[{{Baym} {et~al.}(1969){Baym}, {Pethick}, \&
  {Pines}}]{1969Natur.224..673B}
{Baym}, G., {Pethick}, C., \& {Pines}, D. 1969, \nat, 224, 673

\bibitem[{{Braithwaite}(2009)}]{2009MNRAS.397..763B}
{Braithwaite}, J. 2009, \mnras, 397, 763, 0810.1049

\bibitem[{{Caplan} {et~al.}(2018){Caplan}, {Schneider}, \&
  {Horowitz}}]{2018PhRvL.121m2701C}
{Caplan}, M.~E., {Schneider}, A.~S., \& {Horowitz}, C.~J. 2018, \prl, 121,
  132701, 1807.02557

\bibitem[{{Cardall} {et~al.}(2001){Cardall}, {Prakash}, \&
  {Lattimer}}]{2001ApJ...554..322C}
{Cardall}, C.~Y., {Prakash}, M., \& {Lattimer}, J.~M. 2001, \apj, 554, 322,
  astro-ph/0011148

\bibitem[{{Carreau} {et~al.}(2020){Carreau}, {Gulminelli}, {Chamel}, {Fantina},
  \& {Pearson}}]{2020A&A...635A..84C}
{Carreau}, T., {Gulminelli}, F., {Chamel}, N., {Fantina}, A.~F., \& {Pearson},
  J.~M. 2020, \aap, 635, A84, 1912.01265

\bibitem[{{Chandrasekhar}(1962)}]{1962ApPhL...1....7C}
{Chandrasekhar}, B.~S. 1962, Applied Physics Letters, 1, 7

\bibitem[{{Chime/Frb Collaboration} {et~al.}(2020){Chime/Frb Collaboration},
  {Amiri}, {Andersen}, {Bandura}, {Bhardwaj}, {Boyle}, {Brar}, {Chawla},
  {Chen}, {Cliche}, {Cubranic}, {Deng}, {Denman}, {Dobbs}, {Dong}, {Fandino},
  {Fonseca}, {Gaensler}, {Giri}, {Good}, {Halpern}, {Hessels}, {Hill},
  {H{\"o}fer}, {Josephy}, {Kania}, {Karuppusamy}, {Kaspi}, {Keimpema},
  {Kirsten}, {Landecker}, {Lang}, {Leung}, {Li}, {Lin}, {Marcote}, {Masui},
  {McKinven}, {Mena-Parra}, {Merryfield}, {Michilli}, {Milutinovic},
  {Mirhosseini}, {Naidu}, {Newburgh}, {Ng}, {Nimmo}, {Paragi}, {Patel}, {Pen},
  {Pinsonneault-Marotte}, {Pleunis}, {Rafiei-Ravandi}, {Rahman}, {Ransom},
  {Renard}, {Sanghavi}, {Scholz}, {Shaw}, {Shin}, {Siegel}, {Singh}, {Smegal},
  {Smith}, {Stairs}, {Tendulkar}, {Tretyakov}, {Vanderlinde}, {Wang}, {Wang},
  {Wulf}, {Yadav}, \& {Zwaniga}}]{2020Natur.582..351C}
{Chime/Frb Collaboration} {et~al.} 2020, \nat, 582, 351, 2001.10275

\bibitem[{{Clogston}(1962)}]{1962PhRvL...9..266C}
{Clogston}, A.~M. 1962, \prl, 9, 266

\bibitem[{{Cordes}(1993)}]{1993ASPC...36...43C}
{Cordes}, J.~M. 1993, in Astronomical Society of the Pacific Conference Series,
  Vol.~36, Planets Around Pulsars, ed. J.~A. {Phillips}, S.~E. {Thorsett}, \&
  S.~R. {Kulkarni} (Astronomical Society of the Pacific), 43--60

\bibitem[{{Cruces} {et~al.}(2021){Cruces}, {Spitler}, {Scholz}, {Lynch},
  {Seymour}, {Hessels}, {Gouiff{\'e}s}, {Hilmarsson}, {Kramer}, \&
  {Munjal}}]{2021MNRAS.500..448C}
{Cruces}, M. {et~al.} 2021, \mnras, 500, 448, 2008.03461

\bibitem[{{Cutler}(2002)}]{2002PhRvD..66h4025C}
{Cutler}, C. 2002, \prd, 66, 084025, gr-qc/0206051

\bibitem[{{Dong} {et~al.}(2017){Dong}, {Lombardo}, {Zhang}, \&
  {Zuo}}]{2017PAN....80...77D}
{Dong}, J.~M., {Lombardo}, U., {Zhang}, H.~F., \& {Zuo}, W. 2017, Physics of
  Atomic Nuclei, 80, 77

\bibitem[{{Frieben} \& {Rezzolla}(2012)}]{2012MNRAS.427.3406F}
{Frieben}, J., \& {Rezzolla}, L. 2012, \mnras, 427, 3406, 1207.4035

\bibitem[{{Fulde} \& {Ferrell}(1964)}]{1964PhRv..135..550F}
{Fulde}, P., \& {Ferrell}, R.~A. 1964, Physical Review, 135, 550

\bibitem[{{Gezerlis} {et~al.}(2014){Gezerlis}, {Pethick}, \&
  {Schwenk}}]{2014arXiv1406.6109G}
{Gezerlis}, A., {Pethick}, C.~J., \& {Schwenk}, A. 2014, ArXiv e-prints,
  1406.6109

\bibitem[{{Glampedakis} {et~al.}(2011){Glampedakis}, {Jones}, \&
  {Samuelsson}}]{2011MNRAS.413.2021G}
{Glampedakis}, K., {Jones}, D.~I., \& {Samuelsson}, L. 2011, \mnras, 413, 2021,
  1010.1153

\bibitem[{{Glampedakis} \& {Lasky}(2016)}]{2016MNRAS.463.2542G}
{Glampedakis}, K., \& {Lasky}, P.~D. 2016, \mnras, 463, 2542, 1607.05576

\bibitem[{{Goldreich}(1970)}]{1970ApJ...160L..11G}
{Goldreich}, P. 1970, \apjl, 160, L11

\bibitem[{{Goldreich} \& {Sridhar}(1995)}]{1995ApJ...438..763G}
{Goldreich}, P., \& {Sridhar}, S. 1995, \apj, 438, 763

\bibitem[{{Gottfried}(1966)}]{1966qume.book.....G}
{Gottfried}, K. 1966, {Quantum mechanics - Vol.1: Fundamentals} (Reading:W. A.
  Benjamin)

\bibitem[{{Gourgouliatos} \&
  {Cumming}(2014{\natexlab{a}})}]{2014PhRvL.112q1101G}
{Gourgouliatos}, K.~N., \& {Cumming}, A. 2014{\natexlab{a}}, \prl, 112, 171101,
  1311.7345

\bibitem[{{Gourgouliatos} \&
  {Cumming}(2014{\natexlab{b}})}]{2014MNRAS.438.1618G}
------. 2014{\natexlab{b}}, \mnras, 438, 1618, 1311.7004

\bibitem[{{Gourgouliatos} \& {Pons}(2020)}]{2020arXiv200103335G}
{Gourgouliatos}, K.~N., \& {Pons}, J.~A. 2020, arXiv e-prints,
  arXiv:2001.03335, 2001.03335

\bibitem[{{Guo} {et~al.}(2019){Guo}, {Dong}, {Shang}, {Zhang}, {Zuo},
  {Colonna}, \& {Lombardo}}]{2019NuPhA.986...18G}
{Guo}, W., {Dong}, J.~M., {Shang}, X., {Zhang}, H.~F., {Zuo}, W., {Colonna},
  M., \& {Lombardo}, U. 2019, \nphysa, 986, 18, 1810.02709

\bibitem[{{Gusakov} {et~al.}(2017){Gusakov}, {Kantor}, \&
  {Ofengeim}}]{2017PhRvD..96j3012G}
{Gusakov}, M.~E., {Kantor}, E.~M., \& {Ofengeim}, D.~D. 2017, \prd, 96, 103012,
  1705.00508

\bibitem[{{Hashimoto} {et~al.}(1984){Hashimoto}, {Seki}, \&
  {Yamada}}]{1984PThPh..71..320H}
{Hashimoto}, M., {Seki}, H., \& {Yamada}, M. 1984, Progress of Theoretical
  Physics, 71, 320

\bibitem[{{Haskell} \& {Sedrakian}(2018)}]{2018ASSL..457..401H}
{Haskell}, B., \& {Sedrakian}, A. 2018, in Astrophysics and Space Science
  Library, ed. L.~{Rezzolla}, P.~{Pizzochero}, D.~I. {Jones}, N.~{Rea}, \&
  I.~{Vida{\~n}a}, Vol. 457, 401

\bibitem[{{Henriksson} \& {Wasserman}(2013)}]{2013MNRAS.431.2986H}
{Henriksson}, K.~T., \& {Wasserman}, I. 2013, \mnras, 431, 2986, 1212.5842

\bibitem[{{Iroshnikov}(1963)}]{1963AZh....40..742I}
{Iroshnikov}, P.~S. 1963, \azh, 40, 742

\bibitem[{{Jones}(1975)}]{1975Ap&SS..33..215J}
{Jones}, P.~B. 1975, \apss, 33, 215

\bibitem[{{Kinnunen} {et~al.}(2018){Kinnunen}, {Baarsma}, {Martikainen}, \&
  {T{\"o}rm{\"a}}}]{2018RPPh...81d6401K}
{Kinnunen}, J.~J., {Baarsma}, J.~E., {Martikainen}, J.-P., \& {T{\"o}rm{\"a}},
  P. 2018, Reports on Progress in Physics, 81, 046401, 1706.07076

\bibitem[{{Kiuchi} \& {Yoshida}(2008)}]{2008PhRvD..78d4045K}
{Kiuchi}, K., \& {Yoshida}, S. 2008, \prd, 78, 044045, 0802.2983

\bibitem[{{Kraichnan}(1965)}]{1965PhFl....8.1385K}
{Kraichnan}, R.~H. 1965, Physics of Fluids, 8, 1385

\bibitem[{{Lander} \& {Gourgouliatos}(2019)}]{2019MNRAS.486.4130L}
{Lander}, S.~K., \& {Gourgouliatos}, K.~N. 2019, \mnras, 486, 4130, 1902.02121

\bibitem[{{Lander} \& {Jones}(2012)}]{2012MNRAS.424..482L}
{Lander}, S.~K., \& {Jones}, D.~I. 2012, \mnras, 424, 482, 1202.2339

\bibitem[{{Lander} \& {Jones}(2017)}]{2017MNRAS.467.4343L}
------. 2017, \mnras, 467, 4343, 1610.08745

\bibitem[{{Larkin} \& {Ovchinnikov}(1974)}]{1974JETP...38..854L}
{Larkin}, A.~I., \& {Ovchinnikov}, Y.~N. 1974, Soviet Journal of Experimental
  and Theoretical Physics, 38, 854

\bibitem[{{Lasky} \& {Melatos}(2013)}]{2013PhRvD..88j3005L}
{Lasky}, P.~D., \& {Melatos}, A. 2013, \prd, 88, 103005, 1310.7633

\bibitem[{{Lee} {et~al.}(2018){Lee}, {Yoshiike}, \&
  {Tatsumi}}]{2018qcs..confa1006L}
{Lee}, T.-G., {Yoshiike}, R., \& {Tatsumi}, T. 2018, in Quarks and Compact
  Stars 2017 (QCS2017), 011006

\bibitem[{{Levin} {et~al.}(2020){Levin}, {Beloborodov}, \&
  {Bransgrove}}]{2020ApJ...895L..30L}
{Levin}, Y., {Beloborodov}, A.~M., \& {Bransgrove}, A. 2020, \apjl, 895, L30,
  2002.04595

\bibitem[{{Li} {et~al.}(2021){Li}, {Wang}, {Zhu}, {Zhang}, {Zhang}, {Duan},
  {Zhang}, {Feng}, {Tang}, {Chatterjee}, {Cordes}, {Cruces}, {Dai}, {Gajjar},
  {Hobbs}, {Jin}, {Kramer}, {Lorimer}, {Miao}, {Niu}, {Niu}, {Pan}, {Qian},
  {Spitler}, {Werthimer}, {Zhang}, {Wang}, {Xie}, {Yue}, {Zhang}, {Zhi}, \&
  {Zhu}}]{2021arXiv210708205L}
{Li}, D. {et~al.} 2021, arXiv e-prints, arXiv:2107.08205, 2107.08205

\bibitem[{{Li} {et~al.}(2012){Li}, {Spitkovsky}, \&
  {Tchekhovskoy}}]{2012ApJ...746...60L}
{Li}, J., {Spitkovsky}, A., \& {Tchekhovskoy}, A. 2012, \apj, 746, 60,
  1107.0979

\bibitem[{{Li} {et~al.}(2016){Li}, {Levin}, \&
  {Beloborodov}}]{2016ApJ...833..189L}
{Li}, X., {Levin}, Y., \& {Beloborodov}, A.~M. 2016, \apj, 833, 189, 1606.04895

\bibitem[{{Link}(2003)}]{2003PhRvL..91j1101L}
{Link}, B. 2003, Physical Review Letters, 91, 101101, arXiv:astro-ph/0302441

\bibitem[{Link \& Cutler(2002)}]{10.1046/j.1365-8711.2002.05726.x}
Link, B., \& Cutler, C. 2002, Monthly Notices of the Royal Astronomical
  Society, 336, 211,
  https://academic.oup.com/mnras/article-pdf/336/1/211/18417731/336-1-211.pdf

\bibitem[{{Link} {et~al.}(1993){Link}, {Epstein}, \&
  {Baym}}]{1993ApJ...403..285L}
{Link}, B., {Epstein}, R.~I., \& {Baym}, G. 1993, \apj, 403, 285

\bibitem[{{Lorenz} {et~al.}(1993){Lorenz}, {Ravenhall}, \&
  {Pethick}}]{1993PhRvL..70..379L}
{Lorenz}, C.~P., {Ravenhall}, D.~G., \& {Pethick}, C.~J. 1993, \prl, 70, 379

\bibitem[{{Mestel} {et~al.}(1981){Mestel}, {Nittmann}, {Wood}, \&
  {Wright}}]{1981MNRAS.195..979M}
{Mestel}, L., {Nittmann}, J., {Wood}, W.~P., \& {Wright}, G.~A.~E. 1981,
  \mnras, 195, 979

\bibitem[{{Mestel} \& {Takhar}(1972)}]{1972MNRAS.156..419M}
{Mestel}, L., \& {Takhar}, H.~S. 1972, \mnras, 156, 419

\bibitem[{{Mitchell} {et~al.}(2015){Mitchell}, {Braithwaite}, {Reisenegger},
  {Spruit}, {Valdivia}, \& {Langer}}]{2015MNRAS.447.1213M}
{Mitchell}, J.~P., {Braithwaite}, J., {Reisenegger}, A., {Spruit}, H.,
  {Valdivia}, J.~A., \& {Langer}, N. 2015, \mnras, 447, 1213, 1411.7252

\bibitem[{{Mutafchieva} {et~al.}(2019){Mutafchieva}, {Chamel}, {Stoyanov},
  {Pearson}, \& {Mihailov}}]{2019PhRvC..99e5805M}
{Mutafchieva}, Y.~D., {Chamel}, N., {Stoyanov}, Z.~K., {Pearson}, J.~M., \&
  {Mihailov}, L.~M. 2019, \prc, 99, 055805, 1904.05045

\bibitem[{{Nittmann} \& {Wood}(1981)}]{1981MNRAS.196..491N}
{Nittmann}, J., \& {Wood}, W.~P. 1981, \mnras, 196, 491

\bibitem[{{Passamonti} {et~al.}(2017){Passamonti}, {Akg{\"u}n}, {Pons}, \&
  {Miralles}}]{2017MNRAS.465.3416P}
{Passamonti}, A., {Akg{\"u}n}, T., {Pons}, J.~A., \& {Miralles}, J.~A. 2017,
  \mnras, 465, 3416, 1608.00001

\bibitem[{{Pethick} \& {Potekhin}(1998)}]{1998PhLB..427....7P}
{Pethick}, C.~J., \& {Potekhin}, A.~Y. 1998, Physics Letters B, 427, 7,
  astro-ph/9803154

\bibitem[{{Potekhin} \& {Chabrier}(2018)}]{2018A&A...609A..74P}
{Potekhin}, A.~Y., \& {Chabrier}, G. 2018, \aap, 609, A74, 1711.07662

\bibitem[{{Press} {et~al.}(2002){Press}, {Teukolsky}, {Vetterling}, \&
  {Flannery}}]{2002nrca.book.....P}
{Press}, W.~H., {Teukolsky}, S.~A., {Vetterling}, W.~T., \& {Flannery}, B.~P.
  2002, {Numerical recipes in C++ : the art of scientific computing}

\bibitem[{{Rau} \& {Wasserman}(2021)}]{2021MNRAS.tmp.1692R}
{Rau}, P.~B., \& {Wasserman}, I. 2021, \mnras, 2104.08563

\bibitem[{{Rau, P. B.} \& Wasserman(2021)}]{pbrinprep}
{Rau, P. B.}, \& Wasserman, I. 2021, (in preparation)

\bibitem[{{Ravenhall} {et~al.}(1983){Ravenhall}, {Pethick}, \&
  {Wilson}}]{1983PhRvL..50.2066R}
{Ravenhall}, D.~G., {Pethick}, C.~J., \& {Wilson}, J.~R. 1983, \prl, 50, 2066

\bibitem[{{Reisenegger}(2009)}]{2009A&A...499..557R}
{Reisenegger}, A. 2009, \aap, 499, 557, 0809.0361

\bibitem[{{Reisenegger} \& {Goldreich}(1992)}]{1992ApJ...395..240R}
{Reisenegger}, A., \& {Goldreich}, P. 1992, \apj, 395, 240

\bibitem[{{Schneider} {et~al.}(2018){Schneider}, {Caplan}, {Berry}, \&
  {Horowitz}}]{2018PhRvC..98e5801S}
{Schneider}, A.~S., {Caplan}, M.~E., {Berry}, D.~K., \& {Horowitz}, C.~J. 2018,
  \prc, 98, 055801

\bibitem[{{Shaham}(1977)}]{1977ApJ...214..251S}
{Shaham}, J. 1977, \apj, 214, 251

\bibitem[{{Spitzer}(1958)}]{1958IAUS....6..169S}
{Spitzer}, Jr., L. 1958, in IAU Symposium, Vol.~6, Electromagnetic Phenomena in
  Cosmical Physics, ed. {B.~Lehnert}, 169--+

\bibitem[{{Suh} \& {Mathews}(2010)}]{2010ApJ...717..843S}
{Suh}, I.-S., \& {Mathews}, G.~J. 2010, \apj, 717, 843, 1005.2139

\bibitem[{{The CHIME/FRB Collaboration} {et~al.}(2021){The CHIME/FRB
  Collaboration}, {Andersen}, {Bandura}, {Bhardwaj}, {Boyle}, {Brar},
  {Breitman}, {Cassanelli}, {Chatterjee}, {Chawla}, {Cliche}, {Cubranic},
  {Curtin}, {Deng}, {Dobbs}, {Dong}, {Fonseca}, {Gaensler}, {Giri}, {Good},
  {Hill}, {Josephy}, {Kaczmarek}, {Kader}, {Kania}, {Kaspi}, {Leung}, {Li},
  {Lin}, {Masui}, {Mckinven}, {Mena-Parra}, {Merryfield}, {Meyers}, {Michilli},
  {Naidu}, {Newburgh}, {Ng}, {Ordog}, {Patel}, {Pearlman}, {Pen}, {Petroff},
  {Pleunis}, {Rafiei-Ravandi}, {Rahman}, {Ransom}, {Renard}, {Sanghavi},
  {Scholz}, {Shaw}, {Shin}, {Siegel}, {Singh}, {Smith}, {Stairs}, {Tan},
  {Tendulkar}, {Vanderlinde}, {Wiebe}, {Wulf}, \&
  {Zwaniga}}]{2021arXiv210708463T}
{The CHIME/FRB Collaboration} {et~al.} 2021, arXiv e-prints, arXiv:2107.08463,
  2107.08463

\bibitem[{{Thompson} \& {Duncan}(1993)}]{1993ApJ...408..194T}
{Thompson}, C., \& {Duncan}, R.~C. 1993, \apj, 408, 194

\bibitem[{{Wasserman}(2003)}]{2003MNRAS.341.1020W}
{Wasserman}, I. 2003, \mnras, 341, 1020, arXiv:astro-ph/0208378

\bibitem[{{Zanazzi} \& {Lai}(2020)}]{2020ApJ...892L..15Z}
{Zanazzi}, J.~J., \& {Lai}, D. 2020, \apjl, 892, L15, 2002.05752

\bibitem[{{Zhang} {et~al.}(2018){Zhang}, {Gajjar}, {Foster}, {Siemion},
  {Cordes}, {Law}, \& {Wang}}]{2018ApJ...866..149Z}
{Zhang}, Y.~G., {Gajjar}, V., {Foster}, G., {Siemion}, A., {Cordes}, J., {Law},
  C., \& {Wang}, Y. 2018, \apj, 866, 149, 1809.03043

\bibitem[{{Zuo} {et~al.}(2008){Zuo}, {Cui}, {Lombardo}, \&
  {Schulze}}]{2008PhRvC..78a5805Z}
{Zuo}, W., {Cui}, C.~X., {Lombardo}, U., \& {Schulze}, H. 2008, \prc, 78,
  015805

\end{thebibliography}

\end{document}